\documentclass[10pt,twocolumn,letterpaper]{article}

\usepackage{cvpr}
\usepackage{times}
\usepackage{epsfig}
\usepackage{graphicx}
\usepackage{amsmath}
\usepackage{amssymb}

\usepackage{caption}
\usepackage{multirow}
\usepackage{url}
\usepackage{subcaption}

\usepackage[pagebackref=true,breaklinks=true,letterpaper=true,colorlinks,bookmarks=false]{hyperref}

\cvprfinalcopy 

\def\cvprPaperID{7206}

\ifcvprfinal\fi
\begin{document}

\title{IntrA: 3D Intracranial Aneurysm Dataset for Deep Learning}

\author{
Xi Yang\textsuperscript{1} \qquad
Ding Xia\textsuperscript{1,2} \qquad
Taichi Kin\textsuperscript{1} \qquad
Takeo Igarashi\textsuperscript{1}\\
\textsuperscript{1}The University of Tokyo \qquad
\textsuperscript{2}South China University of Technology
}


\twocolumn[{%
\renewcommand\twocolumn[1][]{#1}%
\maketitle
\thispagestyle{empty}
\begin{center}
\vspace{-8mm}
    \centering
    \includegraphics[width=1.0\textwidth]{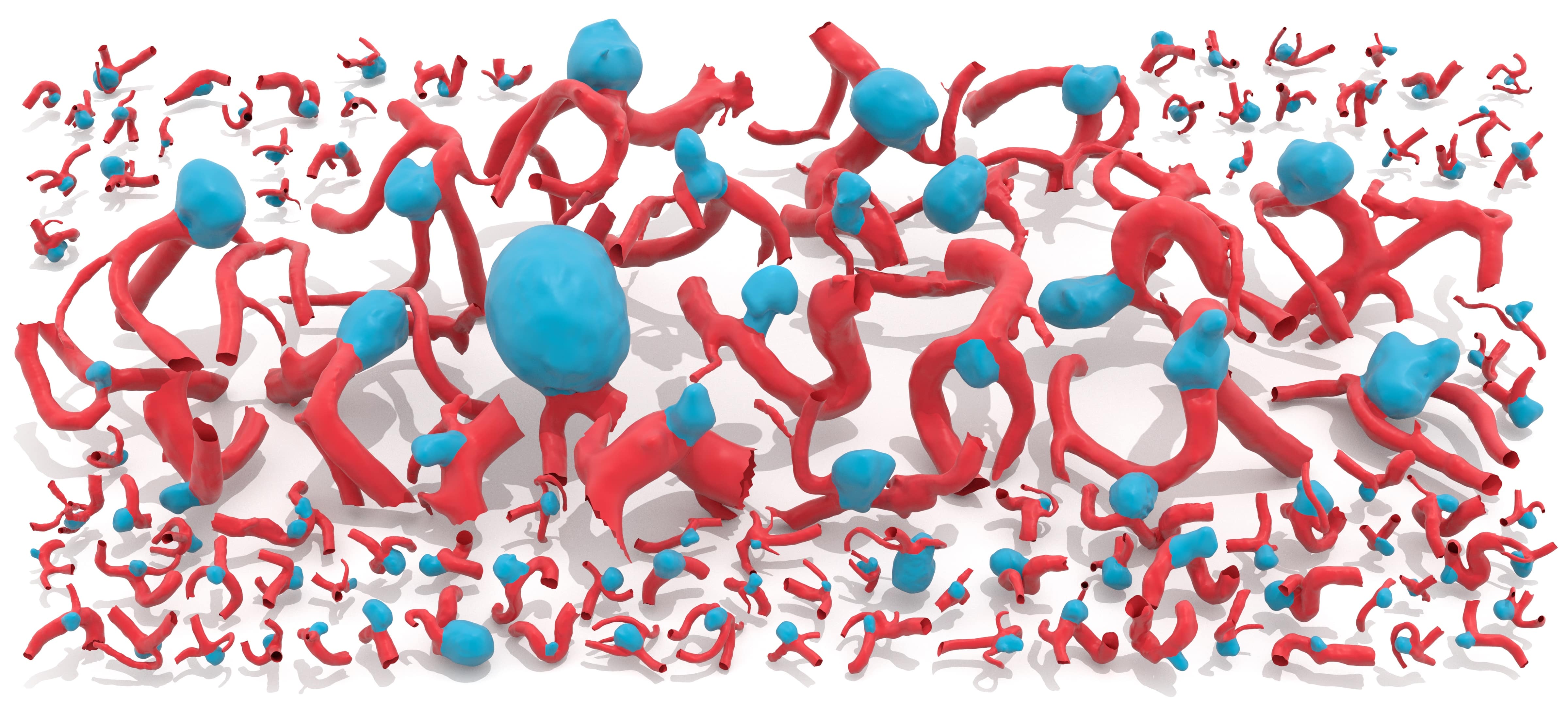}
    \captionof{figure}{3D models of intracranial aneurysm segments with segmentation annotation in our dataset. Hot pink shows the healthy blood vessel part, and aqua shows the aneurysm part for each model.}
    \label{fig:full}
\end{center}%
}]

\begin{abstract}
Medicine is an important application area for deep learning models. Research in this field is a combination of medical expertise and data science knowledge. In this paper, instead of 2D medical images, we introduce an open-access 3D intracranial aneurysm dataset, IntrA, that makes the application of points-based and mesh-based classification and segmentation models available. Our dataset can be used to diagnose intracranial aneurysms and to extract the neck for a clipping operation in medicine and other areas of deep learning, such as normal estimation and surface reconstruction. We provide a large-scale benchmark of classification and part segmentation by testing state-of-the-art networks. We also discuss the performance of each method and demonstrate the challenges of our dataset. The published dataset can be accessed here: \href{https://github.com/intra3d2019/IntrA}{https://github.com/intra3d2019/IntrA}.
\end{abstract}

\vspace{-6mm}
\section{Introduction}
\begin{figure}
\begin{center}
   \includegraphics[width=1.0\linewidth]{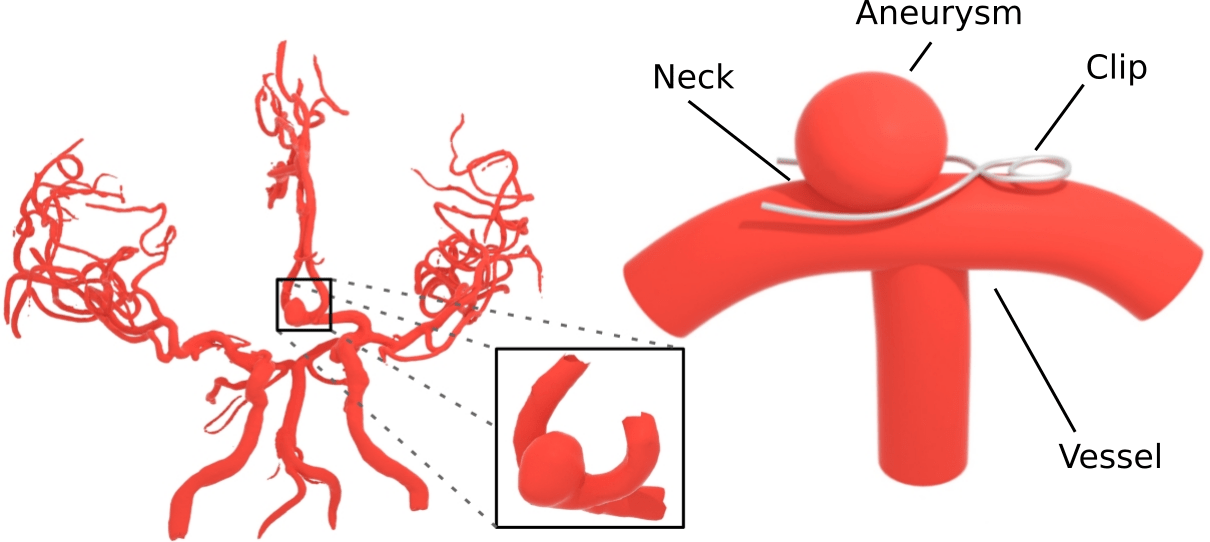}
   \caption{
   The treatment of intracranial aneurysm by clipping.
   }
   \label{fig:clip}
\end{center}
\end{figure}

Intracranial aneurysm is a life-threatening disease, and its surgical treatments are complicated. Timely diagnosis and preoperative examination are necessary to formulate the treatment strategies and surgical approaches. Currently, the primary treatment method is clipping the neck of an aneurysm to prevent it from rupturing, as shown in Figure~\ref{fig:clip}. The decisions of the position and posture of the clip are still highly dependent on “clinical judgment” based on the experience of physicians. In the surgery support system of intracranial aneurysms simulating real-life neurosurgery and teach neurosurgical residents~\cite{alaraj2015virtual}, the accuracy of aneurysm segmentation is the most crucial part because it is used to extract the neck of an aneurysm, that is, the boundary line of the aneurysm.

Based on 3D surface models, the diagnosis of an aneurysm can be much more accurate than 2D images. The edge of the aneurysm is much clearer for doctors, and the complicated and time-consuming annotation of a mess of 2D images is avoided.
There are many reports of automatic diagnosis and segmentation of aneurysms based on medical images, including intracranial aneurysm (IA) and abdominal aortic aneurysm (AAA)~\cite{lareyre2019fully, lopez20193d, sichtermann2019deep}; however, few reports have been published based on 3D models. This is not only because data collection is inefficient, subjective, and challenging to share in medicine, but also the joint knowledge of computer application science and medical science.

Objects with arbitrary shapes are ubiquitous, and a non-Euclidean manifold reveals more critical information than using Euclidean geometry, like complex typologies of brain tissues in neuroscience\cite{bronstein2017geometric}. However, the study of 2D magnetic resonance angiography (MRA) images confines the selection to 3D neural networks based on pixels and voxels, which also omits the information from manifolds. Therefore, we propose an open-access 3D intracranial aneurysm dataset to solve the above issues and to promote the application of deep learning models in medical science. The points and meshes-based models exhibit excellent generalization abilities for 3D deep learning tasks in our experiments.

Our main contributions are:
\begin{enumerate}
\item We propose an open dataset that consists of 3D aneurysm segments with segmentation annotations, automatically generated blood vessel segments, and complete models of scanned blood vessels of the brains. All annotated aneurysm segments are processed as manifold meshes.  

\item We develop tools to generate 3D blood vessel segments from complete models and to annotate a 3D aneurysm model interactively. The data processing pipeline is also introduced.

\item We evaluate the performance of various state-of-the-art 3D deep learning methods on our dataset to provide benchmarks of classification (diagnose) and segmentation of intracranial aneurysms. Furthermore, we analyze the different features of each method from the results obtained.
\end{enumerate}

\section{Related Work}

\subsection{Datasets}
\textbf{Medical dataset.} 
Large-scale samples are required to surmount the challenges of the complexity and heterogeneity of many diseases, but data collection in medical research is costly, it is unattainable for a single research institute. Therefore, data sharing is critical. Several medical datasets have been published online for collaboration on finding a treatment solution.
For example, 
integrated dataset MedPix~\cite{MedPix}, 
bone X-rays dataset MURA~\cite{rajpurkar2017mura},
brain neuroimaging dataset~\cite{fotenos2005normative}, 
Medical Segmentation Decathlon~\cite{DBLP:journals/corr/abs-1902-09063},
Harvard GSP~\cite{DVN/25833_2014}, 
and SCR database~\cite{van_ginneken:2006-1223}. 
Data collection is also critical for a single category of disease, such as, The Lung Image Database Consortium (LIDC-IDRI)~\cite{LIDC-IDRI}, Indian Diabetic Retinopathy Image Dataset (IDRiD)~\cite{IDRiD}, EyePACS~\cite{IDRiD}, and Autism Brain Imaging Data Exchange (ABIDE)~\cite{ABIDE}.
To date, almost all of them are 2D medical images.

\begin{figure*}
\begin{center}
   \includegraphics[width=1.0\linewidth]{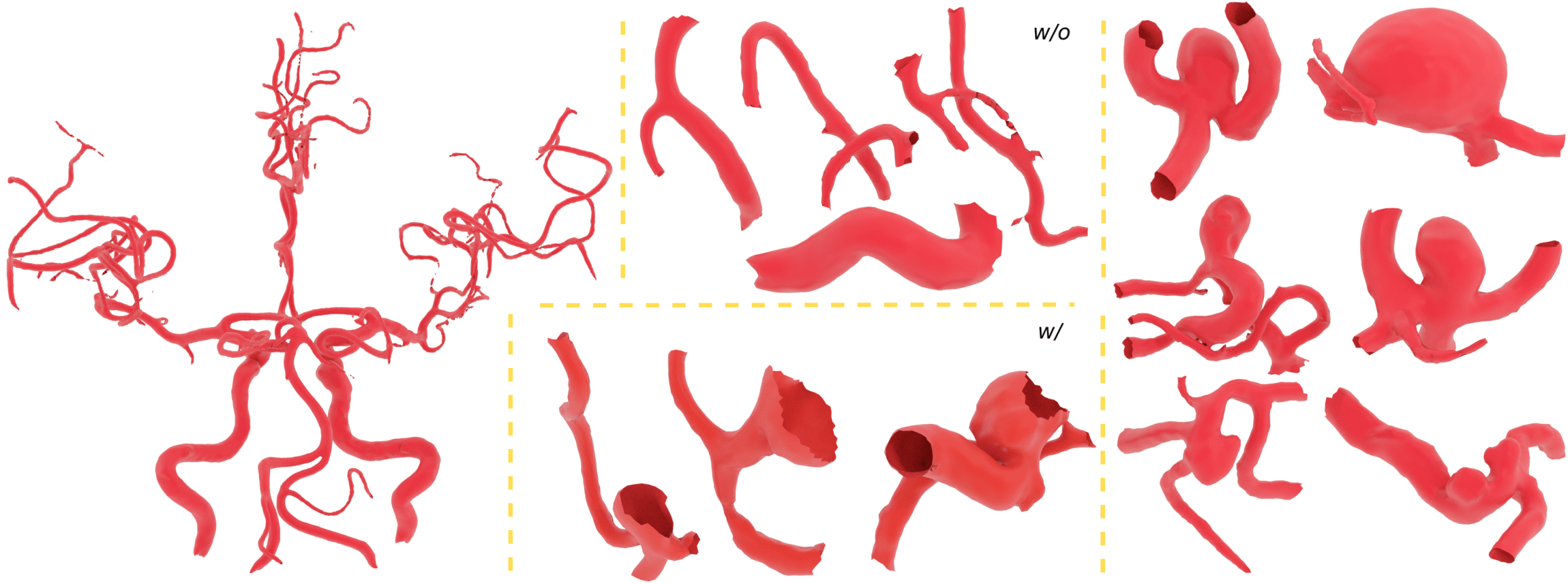}\\
   ~~~~~~~~~~~~~~~~~~~~~ \small{Complete model (103)} ~~~~~~~~~~~~~~~~~~~~~~~~~~~~~~~~ \small{Generated segments (1909)} ~~~~~~~~~~~~~~~~~~~~~~~~~~~~~~~~~ \small{Annotated segments (116)} ~~~~~~~~~
   \caption{There are three types of data in our dataset. The automatically generated blood vessel segments can be with or without aneurysms, and an aneurysm is often partially separated. We see that both aneurysm segments and healthy vessel segments have complex shapes with a different number of branches. For example, in the annotated segments, the aneurysm is huge in the second segment of the first row, but it is tiny in the first segment of the third row. The second segment of the third row has even two aneurysms.}
   \label{fig:data}
\end{center}
\end{figure*}

\textbf{Non-medical 3D dataset.} 
In recent years, 3D model datasets were introduced in the research of computer vision and computer graphics with the development of deep learning algorithms. For instance, CAD model datasets: modelNet~\cite{wu20153d}, shapeNet~\cite{shapenet2015}, COSEG Dataset~\cite{wang2012active}, ABC dataset~\cite{Koch_2019_CVPR}; 3D printing model datasets: Thingi10K~\cite{zhou2016thingi10k}, Human model dataset~\cite{maron2017convolutional}, etc. Various 3D deep learning tasks are widely carried out on these datasets.

\subsection{Methods}
A 3D model has four kinds of representations, projected view, voxel, point cloud, and mesh. The methods based on projected view or voxel are implemented conveniently using similar structures with 2D convolutional neural networks (CNNs). Point cloud or mesh has a more accurate representation of a 3D shape; however, new convolution structures are required. 

\textbf{Projected View.}
Su \etal proposed a multi-view CNN to recognize 3D shapes~\cite{su2015multi}.
Kalogerakis \etal combined image-based fully convolutional networks (FCNs) and surface-based conditional random fields (CRFs) to yield coherent segmentation of 3D shapes~\cite{kalogerakis20173d}.

\textbf{Voxel.}
{\c{C}}i{\c{c}}ek \etal introduced 3D U-Net for volumetric segmentation that learns from sparsely annotated volumetric images~\cite{cciccek20163d}.
Wang \etal presented O-CNN, an Octree-based Convolutional Neural Network (CNN), for 3D shape analysis~\cite{wang2017cnn}.
Graham \etal designed new sparse convolutional operations to process spatially-sparse 3D data, called submanifold sparse convolutional networks (SSCNs)~\cite{graham20183d}.
Wang and Lu proposed VoxSegNet to extract discriminative features encoding detailed information under limited resolution~\cite{wang2019voxsegnet}.
Le and Duan proposed the PointGrid, a 3D convolutional network that is an integration of point and grid~\cite{le2018pointgrid}.

\textbf{Points.}
Qi \etal proposed PointNet, making it is possible to input 3D points directly for neural work~\cite{qi2017pointnet}, 
then they introduced a hierarchical network PointNet++ to learn local features~\cite{qi2017pointnet++}. 
Based on these pioneering works, many new convolution operations were proposed. 
Wu \etal treated convolution kernels as nonlinear functions of the local coordinates of 3D points comprised of weight and density functions, named PointConv~\cite{wu2019pointconv}. 
Li \etal presented PointCNN that can leverage spatially local correlation in data represented densely in grids for feature learning~\cite{li2018pointcnn}. 
Xu \etal designed the filter as a product of a simple step function that captures local geodesic information and a Taylor polynomial, named SpiderCNN~\cite{xu2018spidercnn}.
Moreover,
the SO-Net models the spatial distribution of point cloud by building a Self-Organizing Map (SOM)~\cite{li2018so}.
Su \etal presented SPLATNet for processing point clouds that directly operates on a collection of points represented as a sparse set of samples in a high-dimensional lattice~\cite{su2018splatnet}.
Zhao \etal proposed 3D point-capsule networks~\cite{zhao20193d}.
Wang \etal proposed dynamic graph neural network (DGCNN)~\cite{wang2019dynamic}.
Yang \etal developed Point Attention Transformers (PATs) to process the point clouds~\cite{yang2019modeling}.
Thomas \etal presented a new design of point convolution, called Kernel Point Convolution1
(KPConv)~\cite{thomas2019kpconv}.
Liu \etal proposed a dynamic points agglomeration module to construct an efficient hierarchical point sets learning architecture.~\cite{Liu_2019_ICCV}.

\textbf{Mesh.} Maron \etal applied a convolution operator to sphere-type shapes using a global seamless parameterization to a planar flat-torus~\cite{maron2017convolutional}.
Hanocka \etal utilize the unique properties of the triangular mesh for direct analysis of 3D shapes, named MeshCNN~\cite{hanocka2019meshcnn}.
Feng \etal regard the polygon faces as the unit, split their features into spatial and  structural features called MeshNet~\cite{feng2019meshnet}.



\section{Our Dataset}

\subsection{Data}
Our dataset includes complete models with aneurysms, generated vessel segments, and annotated aneurysm segments, as shown in Figure~\ref{fig:data}. $103$ 3D models of entire brain vessels are collected by reconstructing scanned 2D MRA images of patients. We do not publish the raw 2D MRA images because of medical ethics. $1909$ blood vessel segments are generated automatically from the complete models, including $1694$ healthy vessel segments and $215$ aneurysm segments for diagnosis. An aneurysm can be divided into segments that can verify the automatic diagnosis. $116$ aneurysm segments are divided and annotated manually by medical experts; the scale of each aneurysm segment is based on the need for a preoperative examination. The details are described in the next section. Furthermore, geodesic distance matrices are computed and included for each annotated 3D segment, because the expression of the geodesic distance is more accurate than Euclidean distance according to the shape of vessels. The matrix is saved as ${N}\times{N}$ for a model with $N$ points, shortening the training computation time.

Our data have several characteristics common to medical data: 1) Small but diverse. The amount of data is not so large compared to other released CAD model datasets; however, it includes diverse shapes and scales of intracranial aneurysms as well as different amounts of vessel branches. 2) Unbalanced. The number of points of aneurysms and healthy vessel parts is imbalanced based on the shape of aneurysms. The number of 3D aneurysm segments and healthy vessel segments are not equal because the aneurysms are usually much smaller than the entire brain.

\begin{figure}[t]
\begin{center}
   \includegraphics[width=0.5\linewidth]{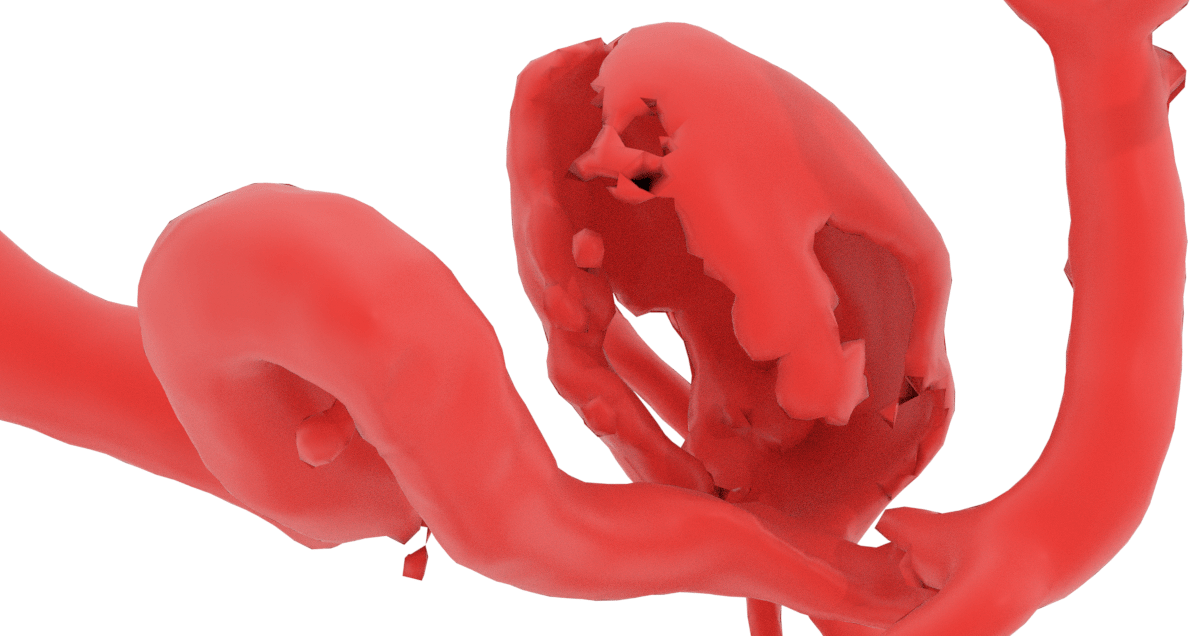}~
   \includegraphics[width=0.5\linewidth]{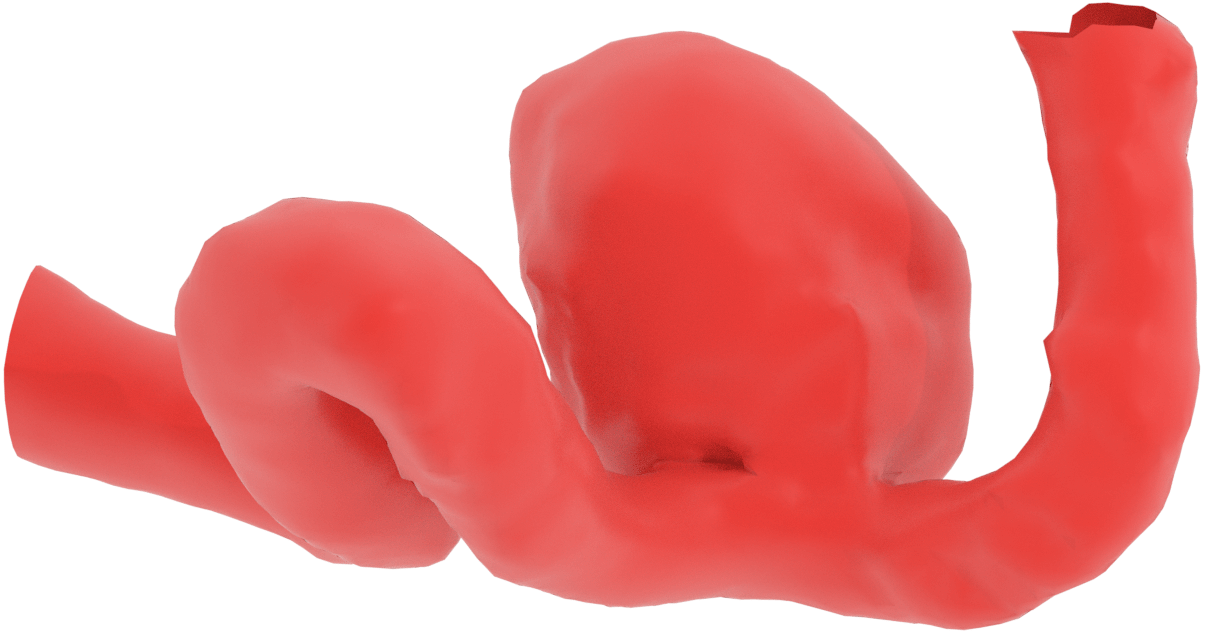}
   \caption{An intracranial aneurysm is restored from incomplete scanned data by neurosurgeons.}
   \label{fig:restore}
\end{center}
\end{figure}

\textbf{Challenge.}
Experts collected out dataset instead of regular people. Intact 3D models have to be restored from reconstructed data manually, as shown in Figure~\ref{fig:restore}. Besides,  the annotation of the neck of aneurysms necks requires years of clinical experience in complex situations. Also, the 3D models are not manifold. We clean the surface meshes to create an ideal dataset for algorithm research.

\begin{figure*}[t]
\begin{center}
   \includegraphics[width=0.345\linewidth]{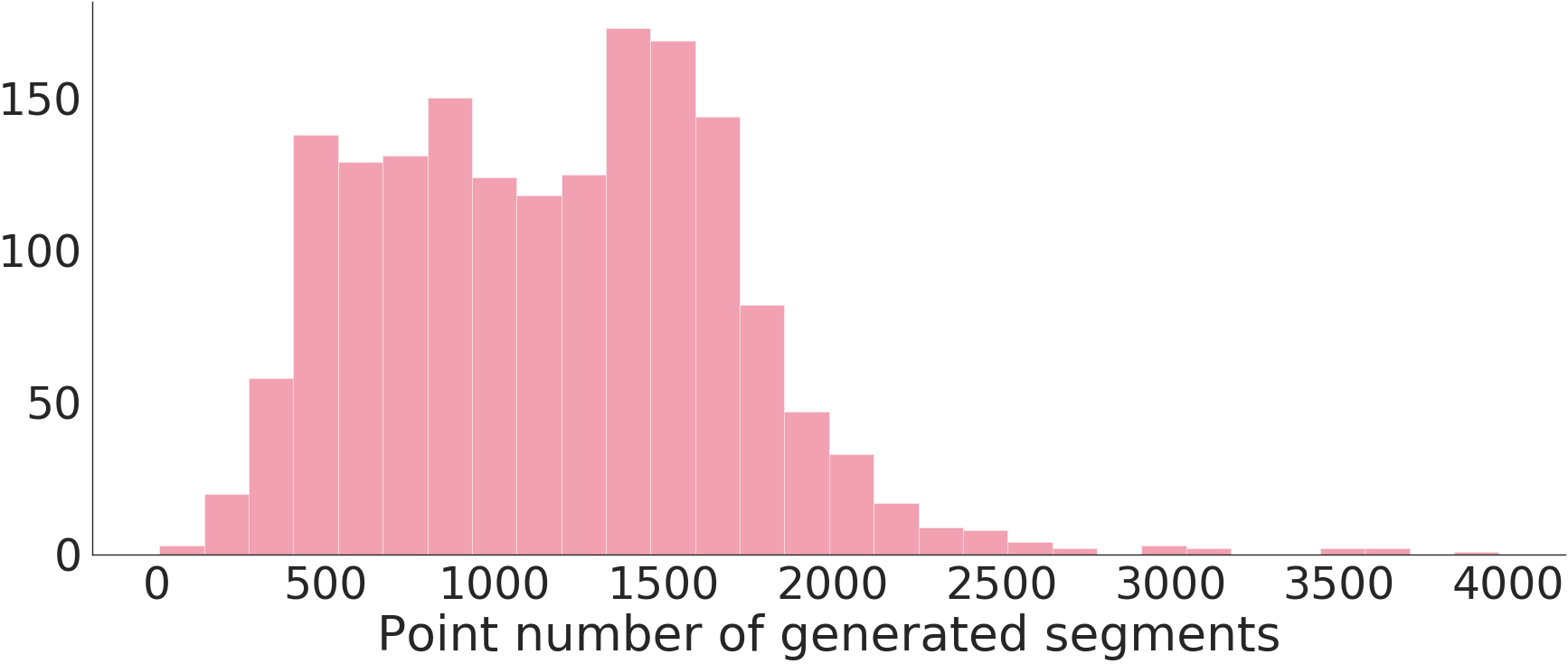}
   \includegraphics[width=0.322\linewidth]{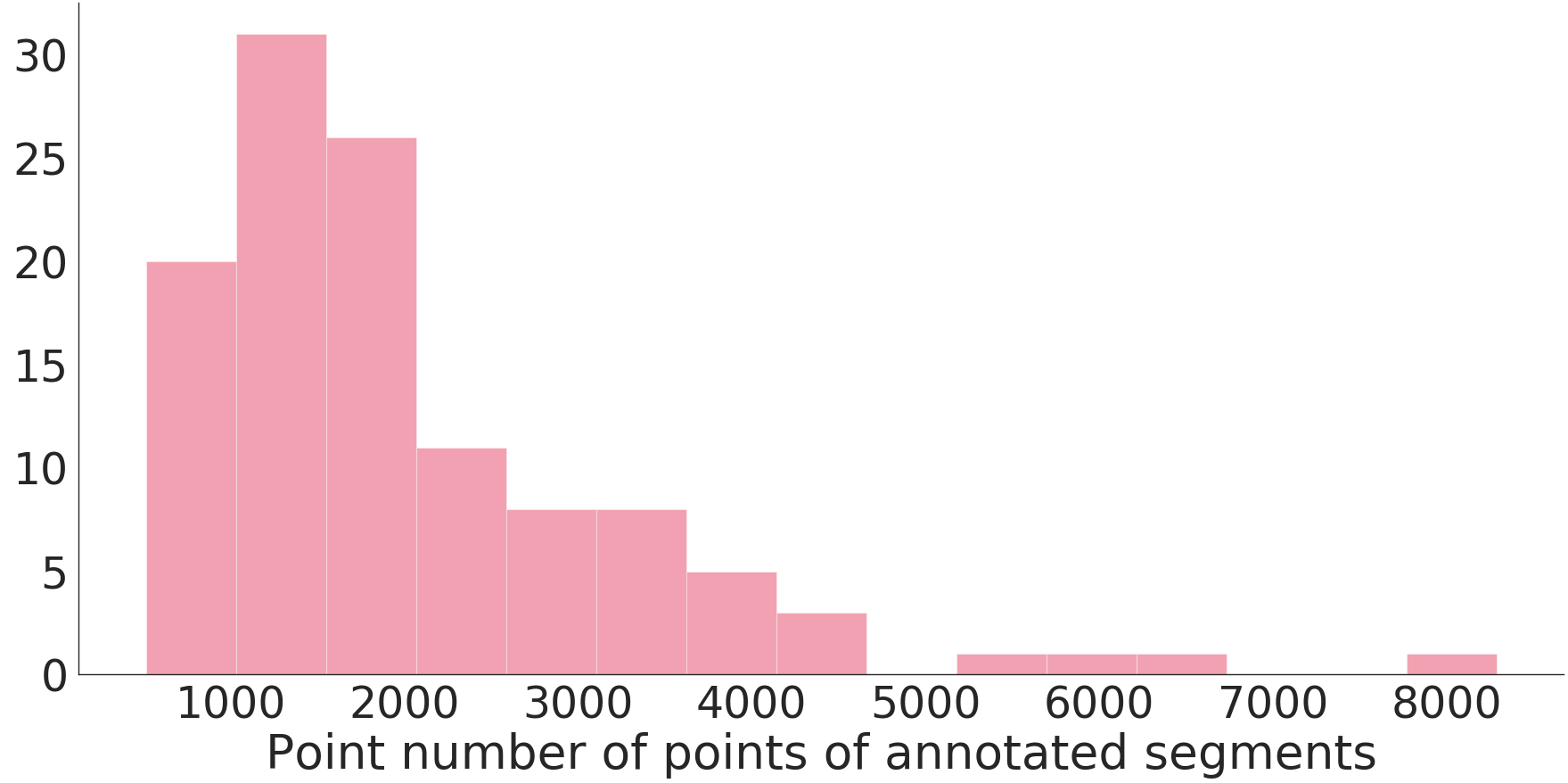}
   \includegraphics[width=0.322\linewidth]{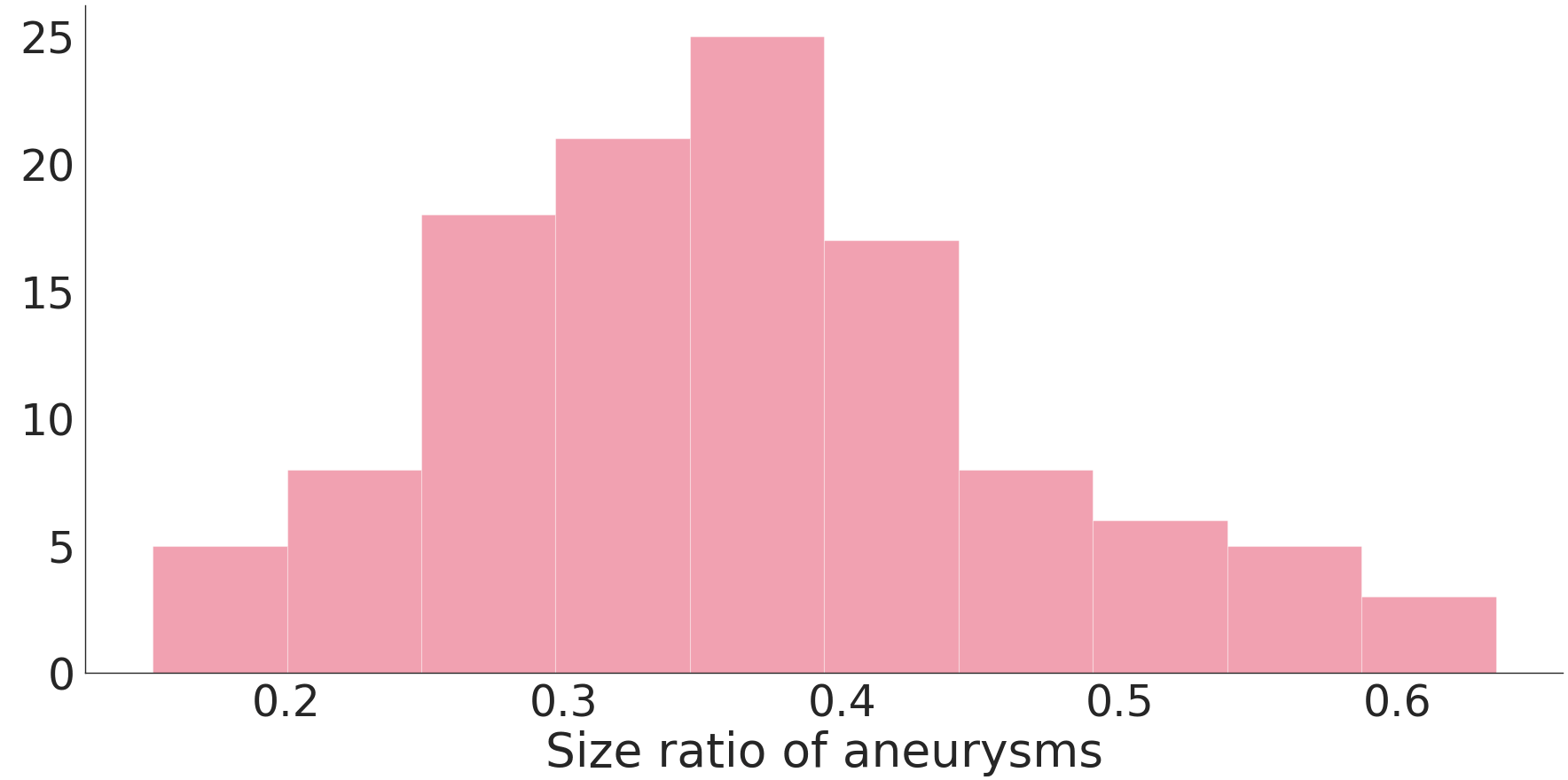}
   \caption{Statistics of annotated models. }
   \label{fig:statistics}
\end{center}
\end{figure*}

\textbf{Statistics and analysis.}
The statistics of our dataset are shown in Figure~\ref{fig:statistics}. We count the number points in each segment to some extent express the difference of the shapes, since the points are mostly uniformly distributed on the surface. The number of points generated in 1909 segments is approximately $500$ to $1700$ at Geodesic distance $30$. 
Our dataset includes all types of intracranial aneurysms in medicine: bifurcation type, trunk type, blister type, and combined type. The shapes of aneurysm are diverse in our dataset both in geometry and topology; six aneurysm segments selected are shown in the right of Figure~\ref{fig:data}. Besides, we calculated the size of an aneurysm as the ratio between the diagonal distance of the global segment and the aneurysm part instead of the real size of the parent vessel or the aneurysm.


\subsection{Tools}
We developed annotation tools and segment generation tools to assist in constructing our dataset. 

\begin{figure}[th]
\begin{center}
   \includegraphics[width=0.95\linewidth]{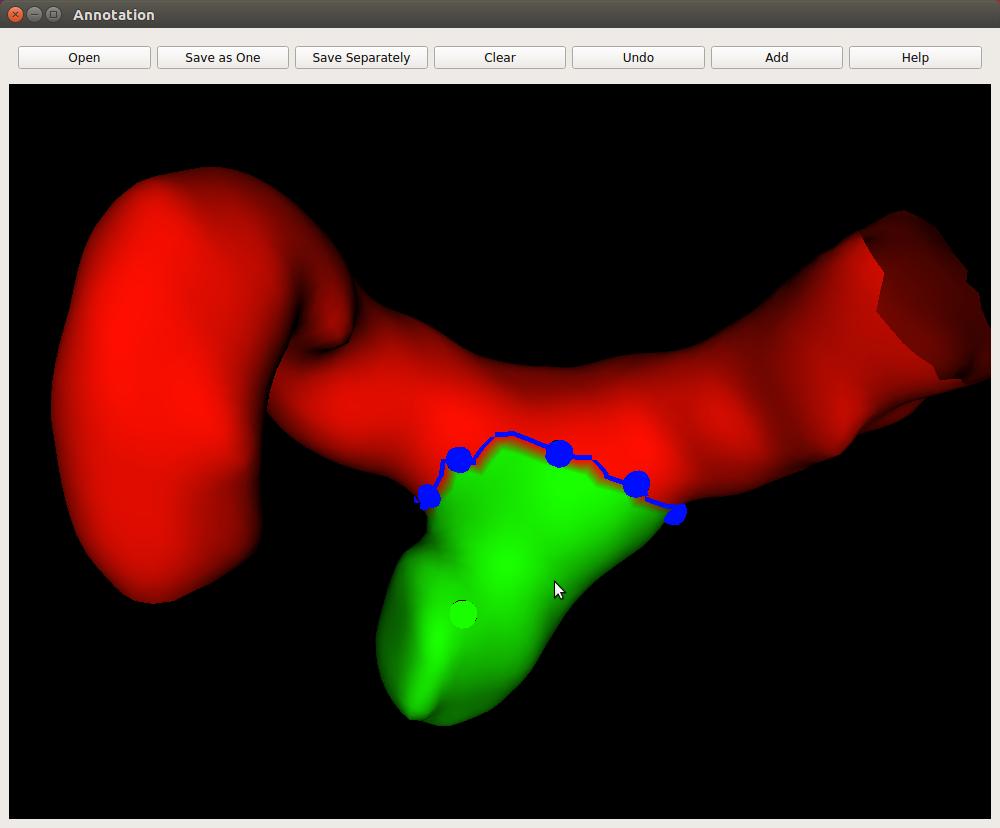} \\
   \includegraphics[width=0.46\linewidth]{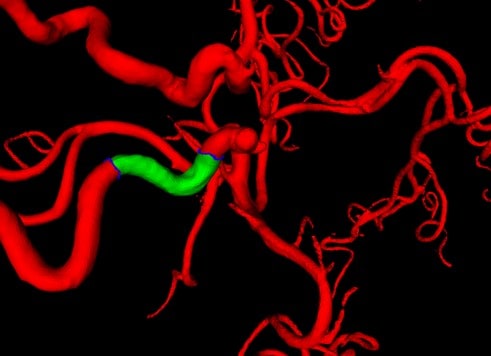}
   \includegraphics[width=0.47\linewidth]{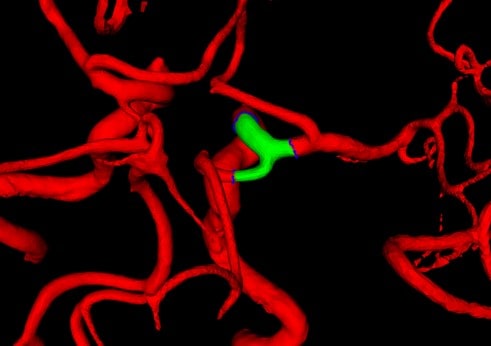}
   \caption{The top figure shows the UI of our annotation tool. The points the user clicked are shown as blue to decide the boundary of the aneurysm, then a random point (yellow) is selected to annotate the aneurysm part (green). The two bottom figures demonstrate the results of multiple boundary lines.}
   \label{fig:ann_tool}
   \vspace{-6mm}
\end{center}
\end{figure}

\textbf{Annotation.}  
Users draw an intended boundary by clicking several points. The connection between two points is determined by the shortest path. After users create a closed boundary line, they annotate the aneurysm part by selecting a point inside of it. The enclosed area is calculated automatically by propagation from the point to the boundary line along with surface meshes. With the support of multiple boundary lines, the annotation tool also can be used for separating both the aneurysm part and the aneurysm segment manually, as shown in Figure~\ref{fig:ann_tool}.

\textbf{Vessel segment generation.} Vessel segments are generated by randomly picking points from the complete models and selecting the neighbor area whose geodesic distance along the vessel is smaller than a threshold. We also manually select points for increasing the number of segments with an aneurysm. To construct an ideal dataset, few data which are ambiguous or only include trivial components are removed by using our visualization tool.

\subsection{Processing pipeline}

\begin{figure*}[th]
\begin{center}
   \includegraphics[width=1.0\linewidth]{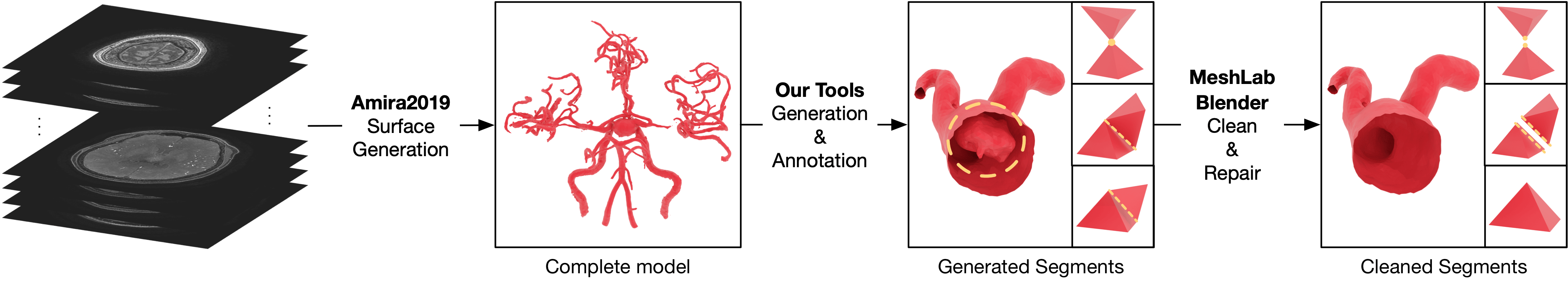}
   \caption{
   Pipeline of data processing. We generate 3D surface models of brain blood vessels from MRA images. Then, the segments are generated and annotated. Data clean and re-meshing are conducted as the requirement of mesh-based methods.}
   \label{fig:pipeline}
\end{center}
\end{figure*}

The processing pipeline is shown in Figure~\ref{fig:pipeline}, more details are described in the supplementary material.

\textbf{3D reconstruction and restore.} Our data are acquired by Time-Of-Flight Magnetic Resonance Angiography (TOF-MRA) of human brain. Using the single threshold method~\cite{hoogeveen1998limits}, each complete 3D model is reconstructed from ${512}\times{512}\times{180}\sim{300}$ 2D images sliced by ${0.469}\times{0.469}\times{1mm}$.
The aneurysm segments are separated and restored interactively using the multi-threshold method~\cite{kin2012new} by two neurosurgeons, then processed by Gaussian smoothing. This image processing is conducted in life sciences software, Amira 2019 (Thermo Fisher Scientific, MA, USA). It takes about 50 workdays in total.

\textbf{Generation and annotation.} By using our generation and annotation tools, blood vessel segments are obtained and classified. The segmentation annotation of aneurysm segments is also finished. A neurosurgeon completed it in 8 hours.

\textbf{Data clean and re-meshing.} The reconstructed 3D models are noisy and not manifold. 
Huang \etal~\cite{huang2018robust} described an algorithm to generate a manifold surface for 3D models; however, this method does not remove isolated components and significantly changes the shape of the model.
Therefore, we use filter tools in MeshLab to remove duplicate faces and vertices, and separate pieces in the data manually, which ensure that the models do not have non-manifold edges. MeshLab also generates the normal vector at each point. The geodesic matrix is computed by solving the heat equation on the surface using a fast approximate geodesic distance method by~\cite{crane2013geodesics}.

\subsection {Supported Studies}

\textbf{Diagnose (Classification).} The diagnosis of an aneurysm can be considered as a classification problem of aneurysms and healthy vessel segments. From a 3D brain model of a patient, vessel segments are generated by our tools; then, the diagnosis is completed by classifying the segments with aneurysms.

\textbf{Part segmentation.} Our annotated 3D models present a precise boundary of each aneurysm to support segmentation research. The data is easy to convert to any 3D representation of various deep learning algorithms.

\textbf{Rule-based algorithms.} Besides, algorithms based on rules are proposed for either aneurysm in the brain or abdominal aortic aneurysm~\cite{dakua2018pca, law2007vessel, salahat2017segmentation}. The accuracy and generalization of the designed rules should be verified by large data.

\textbf{Others.} Because we provide the information about the normal vector and geodesic distance in our dataset, other researches on 3D models are also supported, \eg normal estimation~\cite{boulch2016deep}, geodesic estimation~\cite{he2019geonet}, 3D surface reconstruction~\cite{dai2017shape, park2019deepsdf}, and etc.

\section{Benchmark}
We selected state-of-the-art methods as the benchmarks of classification and segmentation of our dataset. We implemented dataset interfaces to the original implementations by the authors and kept the same hyper-parameters and loss functions of models as in the original papers. A detailed explanation of the implementation of each method is described in the supplementary material. We tested these methods by 5-fold cross-validation. The shuffled data was divided into 5 subsamples and was the same for each method. 4 subsamples were used as training data, 1 was for test data. The experiments were carried out on PCs with GeForce RTX 2080 Ti $\times$2, GeForce GTX 1080 Ti $\times$1. The net training time of all methods was over 92 hours.

\subsection{Classification}

6 methods were selected for \textcolor{red}{binary} classification benchmarks, including PointNet~\cite{qi2017pointnet}, PointNet++ (PN++)~\cite{qi2017pointnet++}, PointCNN~\cite{li2018pointcnn}, SpiderCNN~\cite{xu2018spidercnn}, self-organizing network (SO-Net)~\cite{li2018so}, dynamic graph CNN (DGCNN)~\cite{wang2019dynamic}. We combined the generated blood vessel segments and manually segmented aneurysms in total 2025 as the dataset for testing classification accuracy and F1-Score of each method. The experimental results are shown in Table~\ref{tab:c-results}.

PN++ with 1024 sampling points has the highest accuracy of aneurysms, and PointCNN with 2048 sampled points has the greatest accuracy of artery and F1-Score. The accuracy and F1-score of almost all methods showed an increasing tendency as more sampled points were provided. However, SpiderCNN attained the highest aneurysm detection rate and F1-Score at 1024 sampling points. The majority of misclassified 3D models contained small-sized or incomplete aneurysms that are hard to distinguish from healthy blood vessels.

\begin{table}[t]
\begin{center}
\begin{tabular}{ccccc}
\hline
\\ [-2.1ex]
Network & Input & V. (\%). & A. (\%)  & F1-Score \vspace{2pt} \\ 
\hline 
\\ [-2.0ex]
\multirow{3}{*}{PN++} 
& 512 & $98.52$ & $86.69$ & $0.8928$ \\
& 1024 & $98.52$ & $\mathbf{88.51}$ & $0.9029$ \\
& 2048 & $98.76$ & $87.31$ & $0.9016$ \vspace{2pt} \\ 
\\ [-2.7ex]
\hline
\\ [-2.0ex]
\multirow{3}{*}{SpiderCNN}
& 512 & $98.05$ & $84.58$ & $0.8692$  \\
& 1024 & $97.28$ & $87.90$ & $0.8722$  \\
& 2048 & $97.82$ & $84.89$ & $0.8662$  \vspace{2pt} \\ 
\\ [-2.7ex]
\hline
\\ [-2.0ex]
\multirow{3}{*}{SO-Net}
& 512 & $98.76$ & $84.24$ & $0.8840$  \\
& 1024 & $98.88$ & $81.21$ & $0.8684$  \\
& 2048 & $98.88$ & $83.94$ & $0.8850$  \vspace{2pt} \\ 
\\ [-2.7ex]
\hline
\\ [-2.0ex]
\multirow{3}{*}{PointCNN}
& 512 & $98.38$ & $78.25$ & $0.8494$ \\
& 1024 & $98.79$ & $81.28$ & $0.8748$ \\
& 2048 & $\mathbf{98.95}$ & $85.81$ & $\mathbf{0.9044}$ \vspace{2pt} \\ 
\\ [-2.7ex]
\hline
\\ [-2.0ex]
\multirow{3}{*}{DGCNN}
& 512/10 & $95.22$ & $60.73$ & $0.6578$  \\
& 1024/20 & $95.34$ & $72.21$ & $0.7376$  \\
& 2048/40 & $97.93$ & $83.40$ & $0.8594$  \vspace{2pt} \\ 
\\ [-2.7ex]
\hline
\\ [-2.0ex]
\multirow{3}{*}{PointNet}
& 512 & $94.45$ & $67.66$ & $0.6909$ \\
& 1024 & $94.98$ & $64.96$ & $0.6835$ \\
& 2048 & $93.74$ & $69.50$ & $0.6916$ \vspace{2pt} \\ 
\\ [-2.7ex]
\hline

\end{tabular}
\caption{Classification results of each method. The second column shows the number of input points, the additional input $K$ is required for DGCNN. The accuracies of healthy vessel segments (V.) and aneurysm segments (A.) and F1-Score are calculated by the mean value of each fold.}
\label{tab:c-results}
\end{center}
\end{table}

\subsection{Segmentation}
\label{sec:seg}

\begin{figure*}
\begin{center}
   \includegraphics[width=1.0\linewidth]{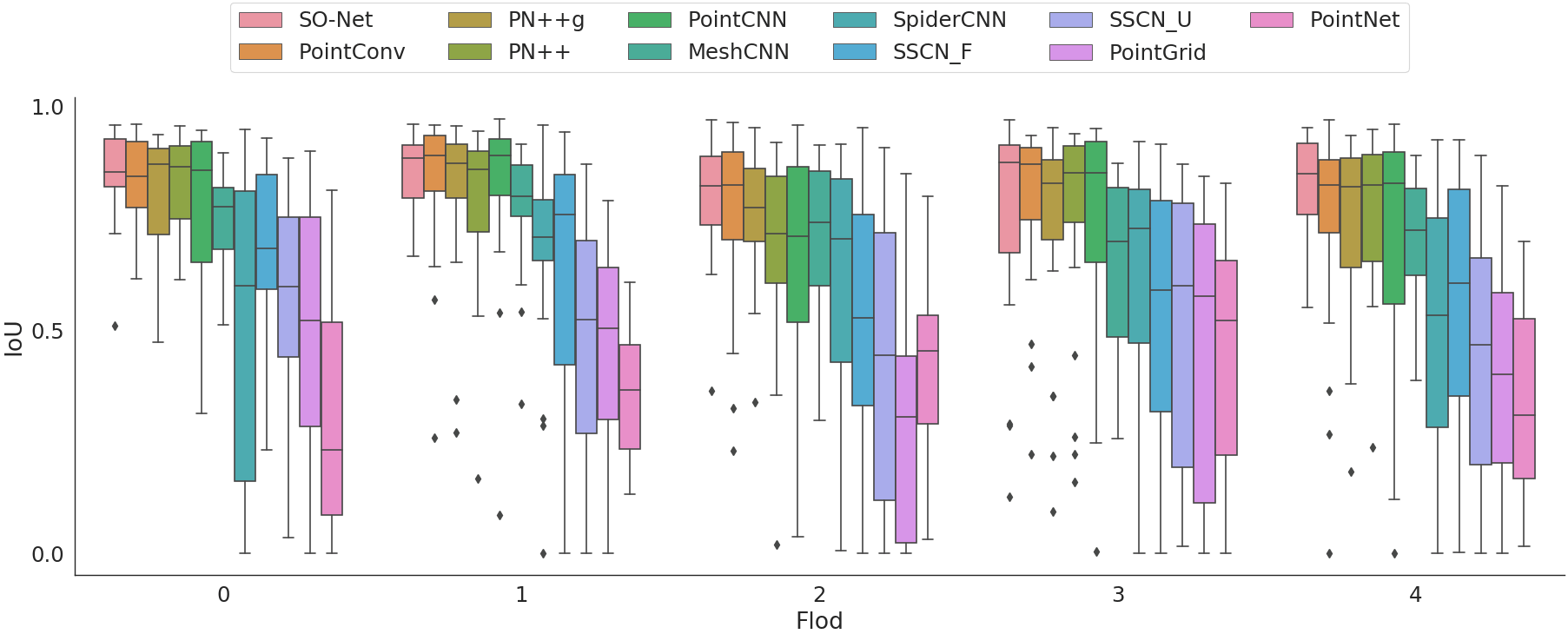}
   \caption{IoU results of the aneurysm part for each fold of networks. These methods are compared with their best performance.}
   \label{fig:iou}
\end{center}
\end{figure*}

We selected 11 networks, PointGrid~\cite{le2018pointgrid}, two kind of submanifold sparse convolutional networks (SSCNs): fully-connected (SSCN-F) and Unet-like (SSCN-U)~\cite{graham20183d} structures, PointNet~\cite{qi2017pointnet}, two kind of PointNet++~\cite{qi2017pointnet++}: input with normal (PN++) and with normal and geodesic distance (PN++g), PointConv~\cite{wu2019pointconv}, PointCNN~\cite{li2018pointcnn}, SpiderCNN~\cite{xu2018spidercnn}, MeshCNN~\cite{hanocka2019meshcnn}, and SO-Net~\cite{li2018so}, to provide segmentation benchmarks. 
116 annotated aneurysm segments were used for evaluating these methods. The test of each subsample was repeated 3 times, and the final results were the mean values of each best result.
We assessed the effects of each method using two indexes: Jaccard Index (JI) and S{\o}rensen-Dice Coefficient (DSC). 
Jaccard Index is also known as Intersection over Union (IoU). 
The results of segmentation are shown in Figure~\ref{fig:iou} and Table~\ref{tab:s-results}. 

\begin{figure}
\vspace{2pt}

\begin{subfigure}{.115\textwidth}
  \centering
  \includegraphics[width=0.95\linewidth]{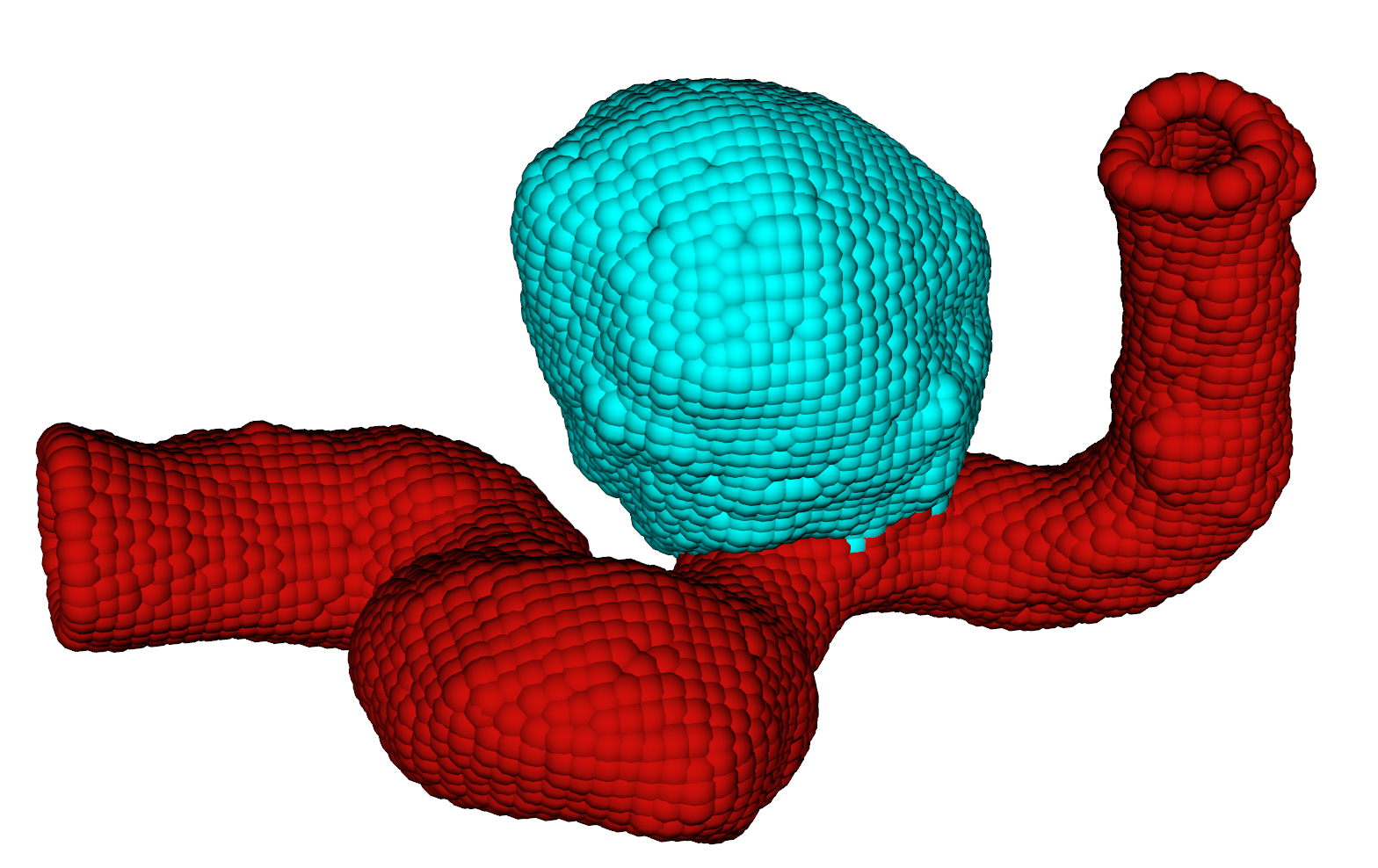}
\end{subfigure}%
\begin{subfigure}{.115\textwidth}
  \centering
  \includegraphics[width=0.9\linewidth]{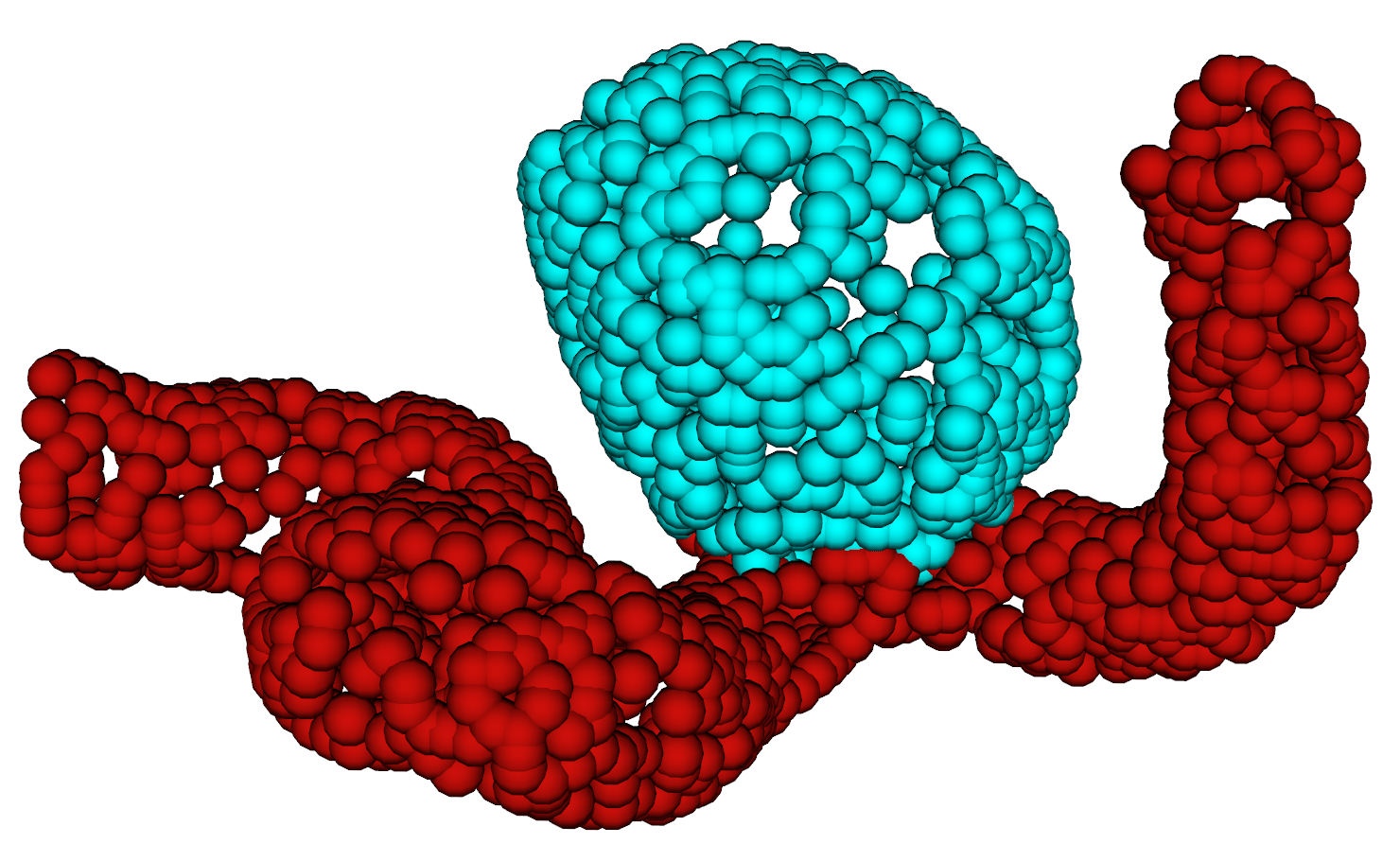}
\end{subfigure}
\begin{subfigure}{.115\textwidth}
  \centering
  \includegraphics[width=0.9\linewidth]{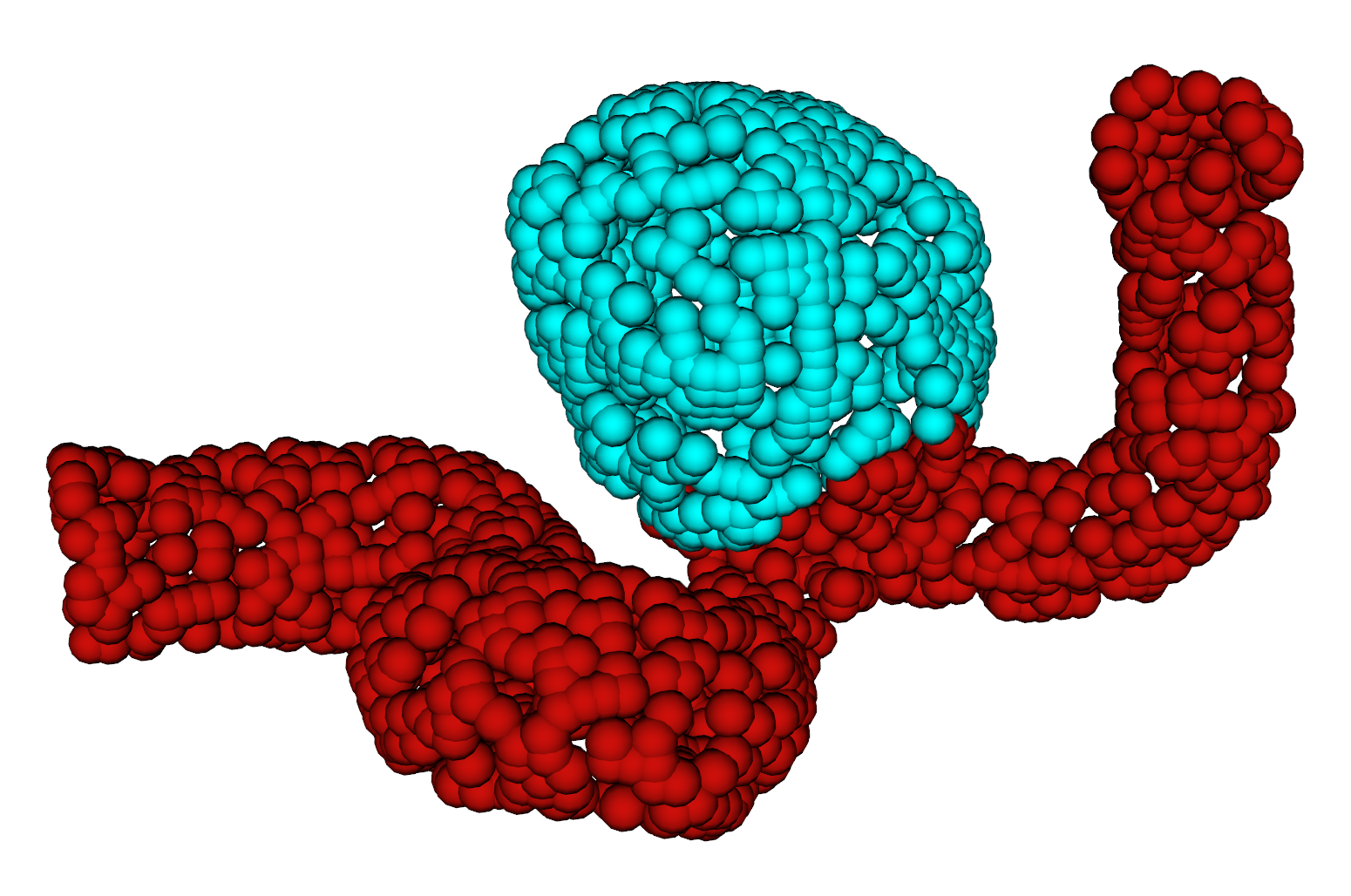}
\end{subfigure}
\begin{subfigure}{.115\textwidth}
  \centering
  \includegraphics[width=0.9\linewidth]{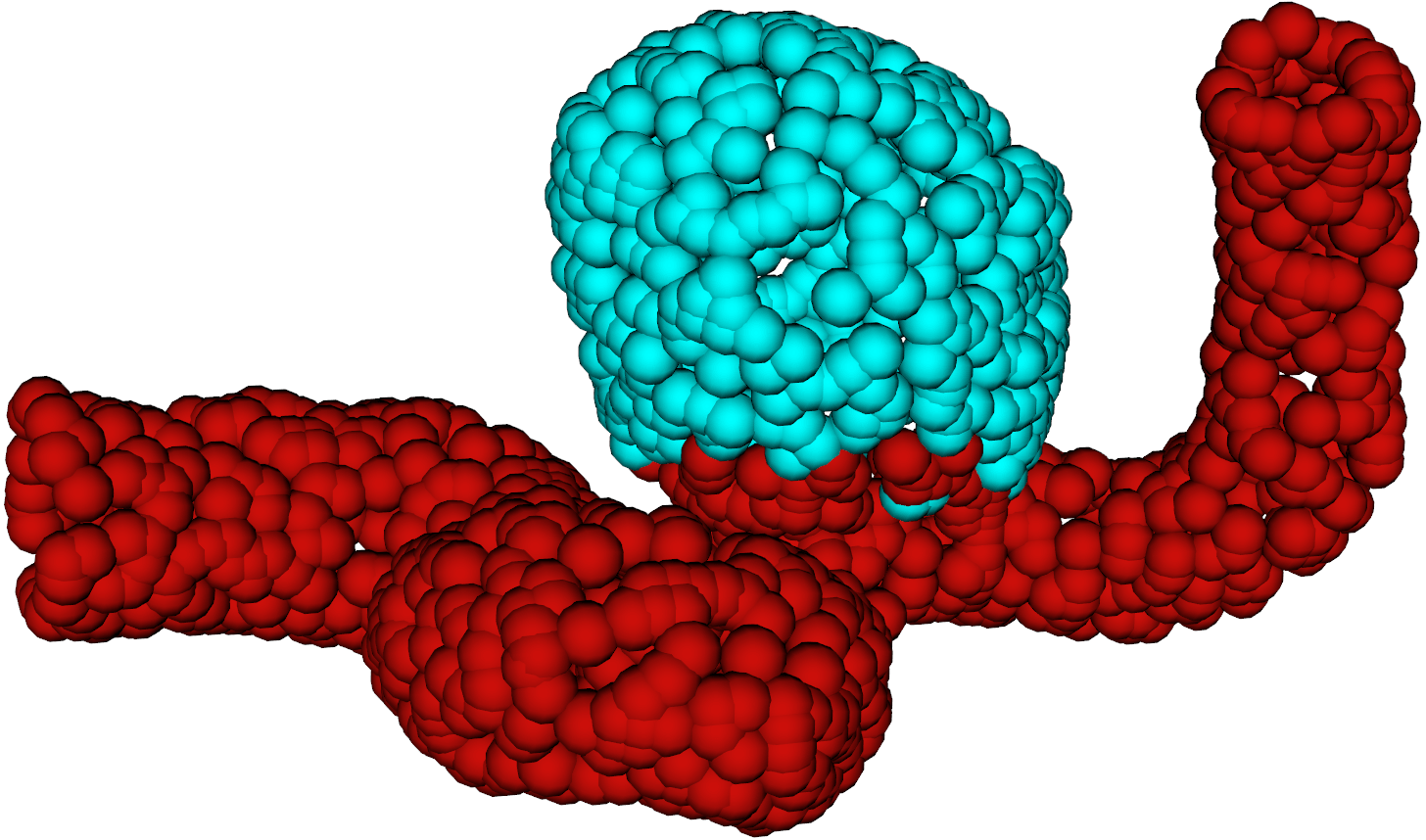}
\end{subfigure}

\vspace{2pt}

\begin{subfigure}{.115\textwidth}
  \centering
  \includegraphics[width=0.9\linewidth]{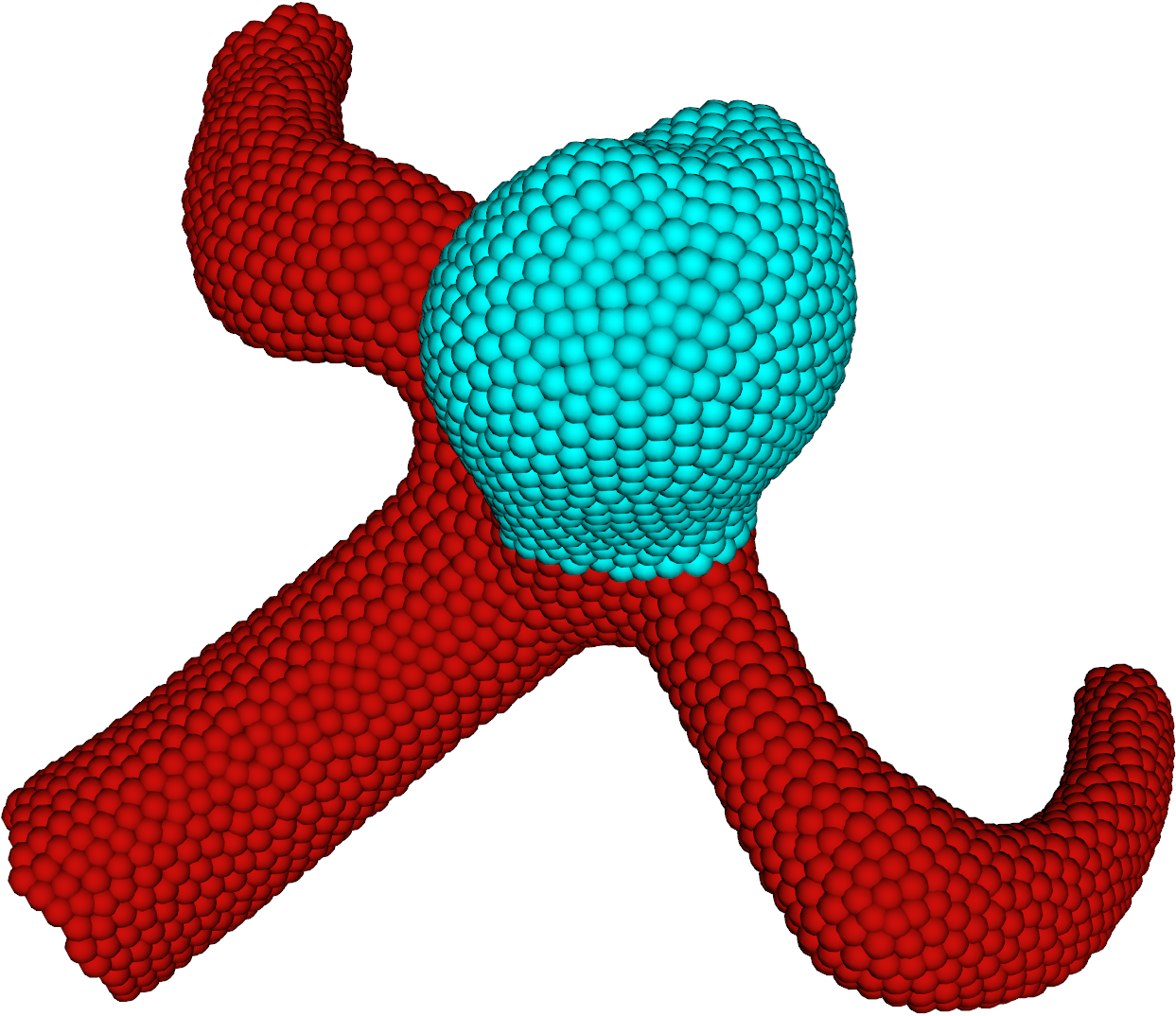}
\end{subfigure}%
\begin{subfigure}{.115\textwidth}
  \centering
  \includegraphics[width=0.9\linewidth]{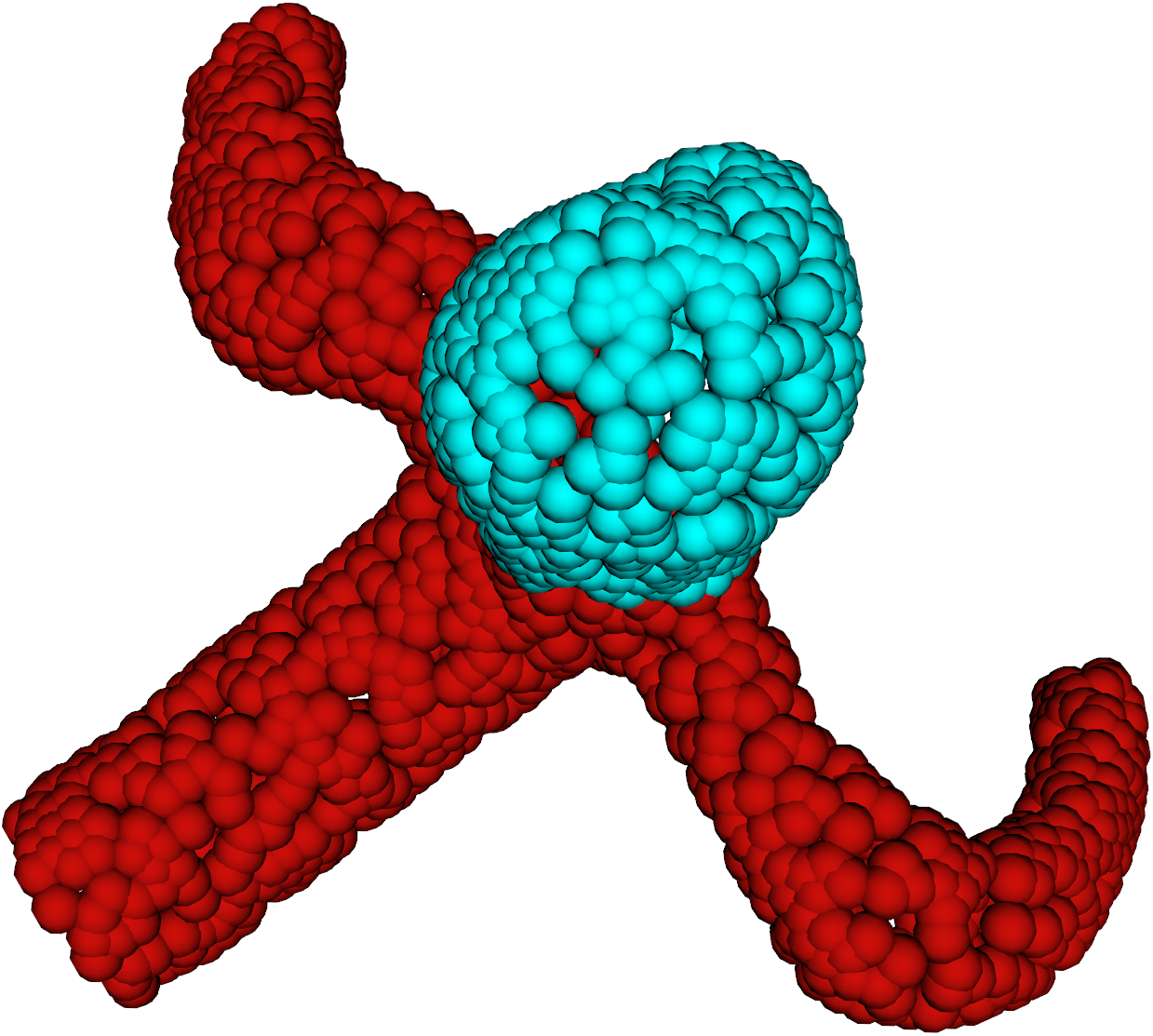}
\end{subfigure}
\begin{subfigure}{.115\textwidth}
  \centering
  \includegraphics[width=0.9\linewidth]{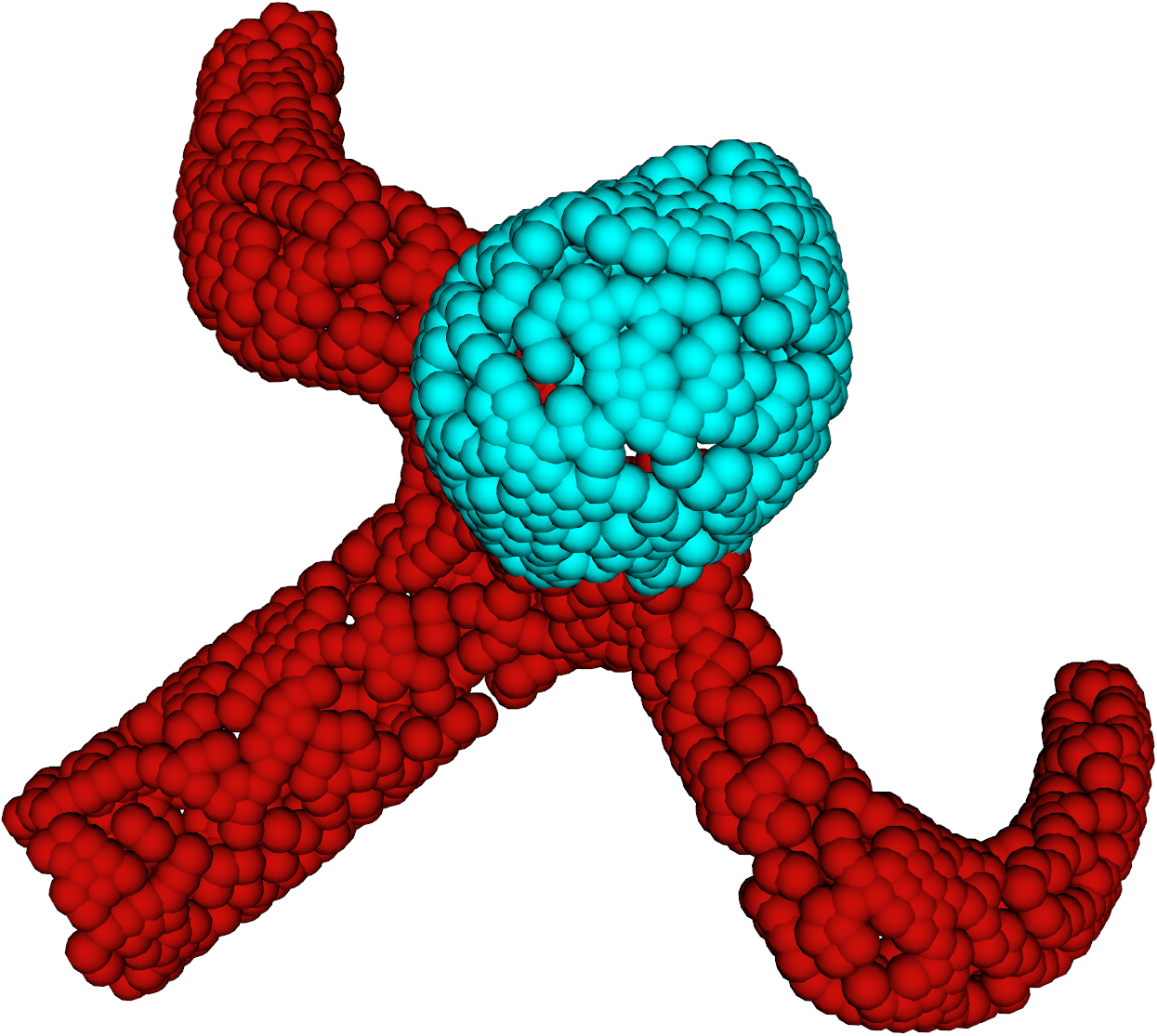}
\end{subfigure}
\begin{subfigure}{.115\textwidth}
  \centering
  \includegraphics[width=0.9\linewidth]{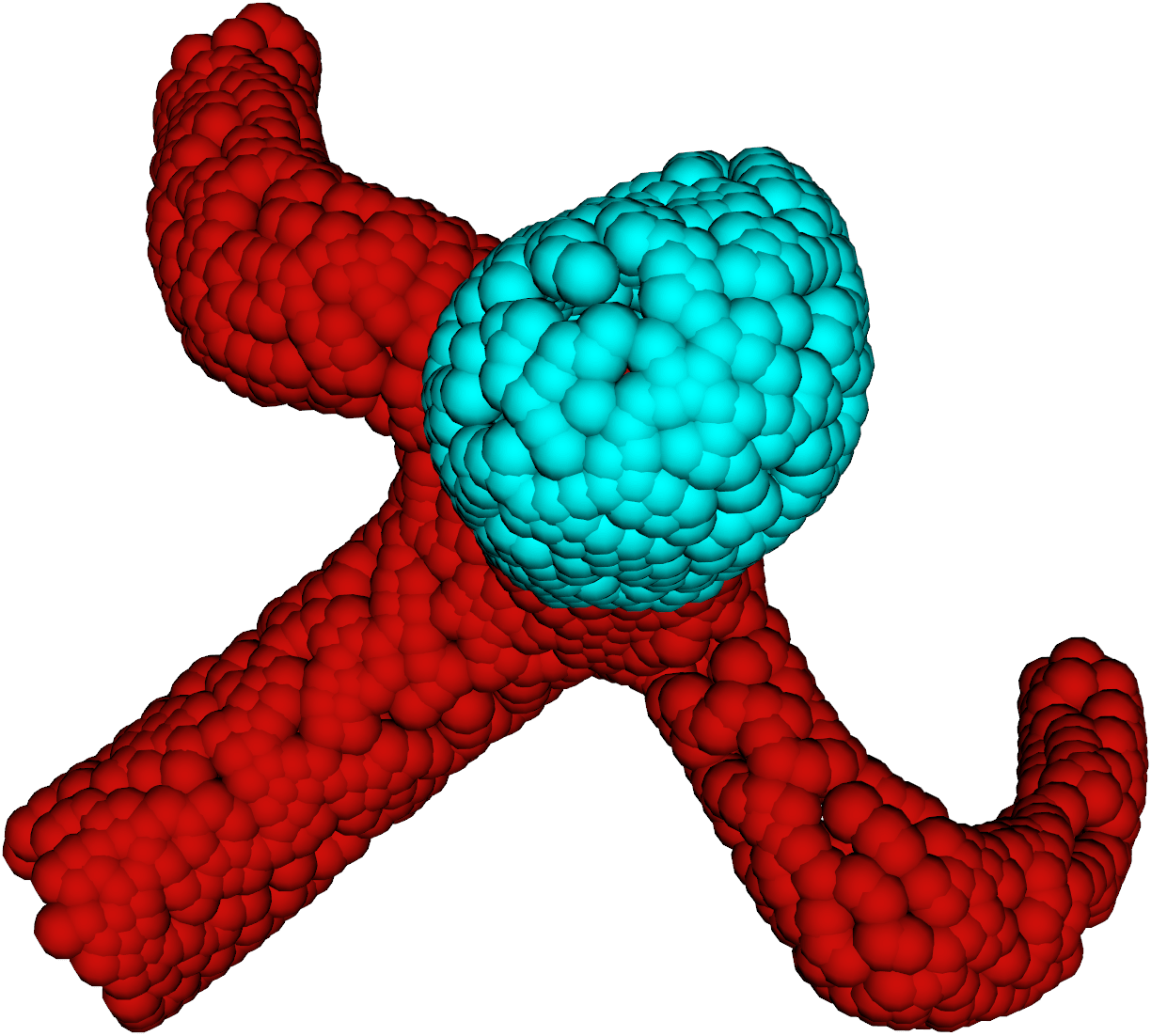}
\end{subfigure}

\vspace{5pt}

\begin{subfigure}{.115\textwidth}
  \centering
  \includegraphics[width=0.9\linewidth]{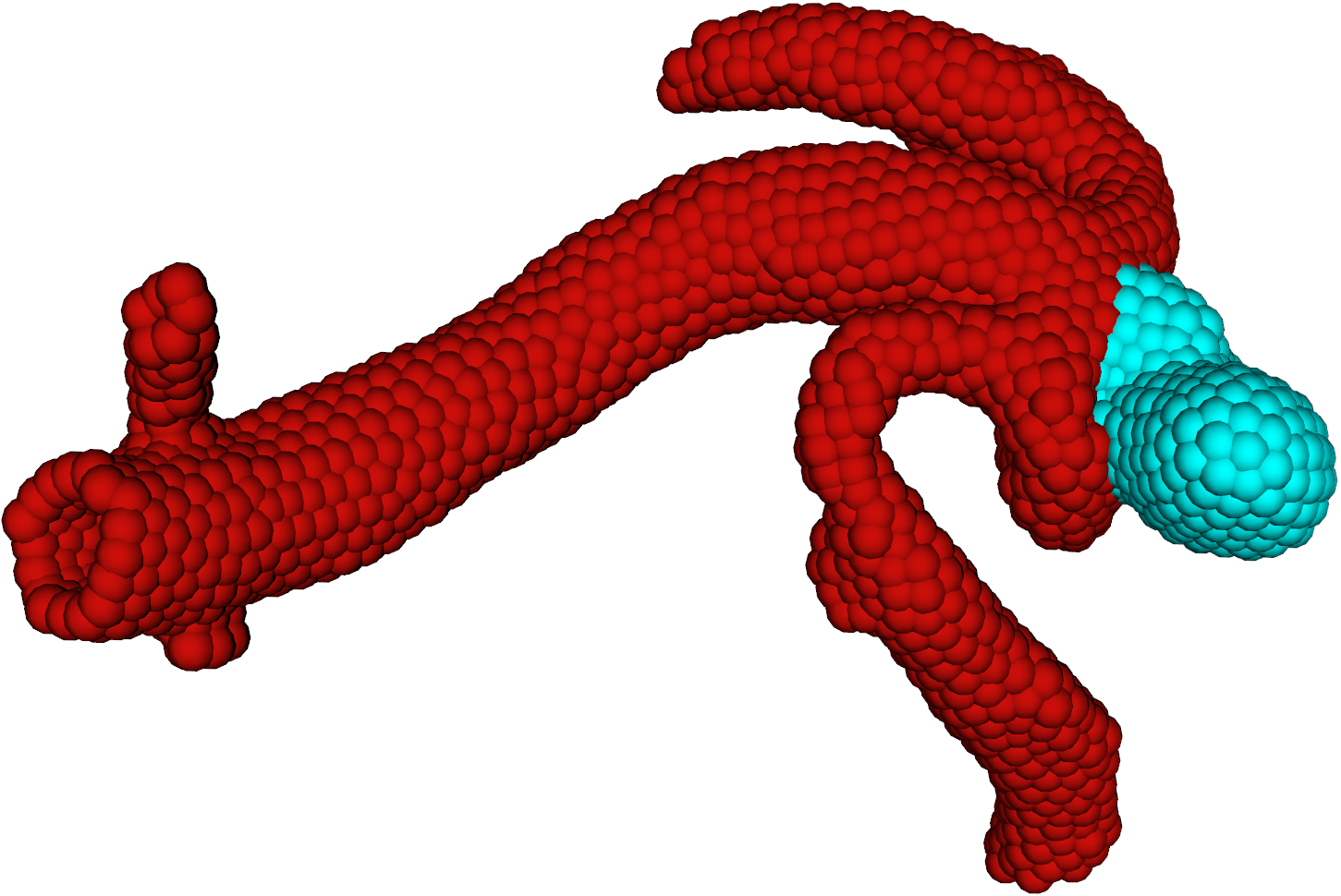}
\end{subfigure}%
\begin{subfigure}{.115\textwidth}
  \centering
  \includegraphics[width=0.9\linewidth]{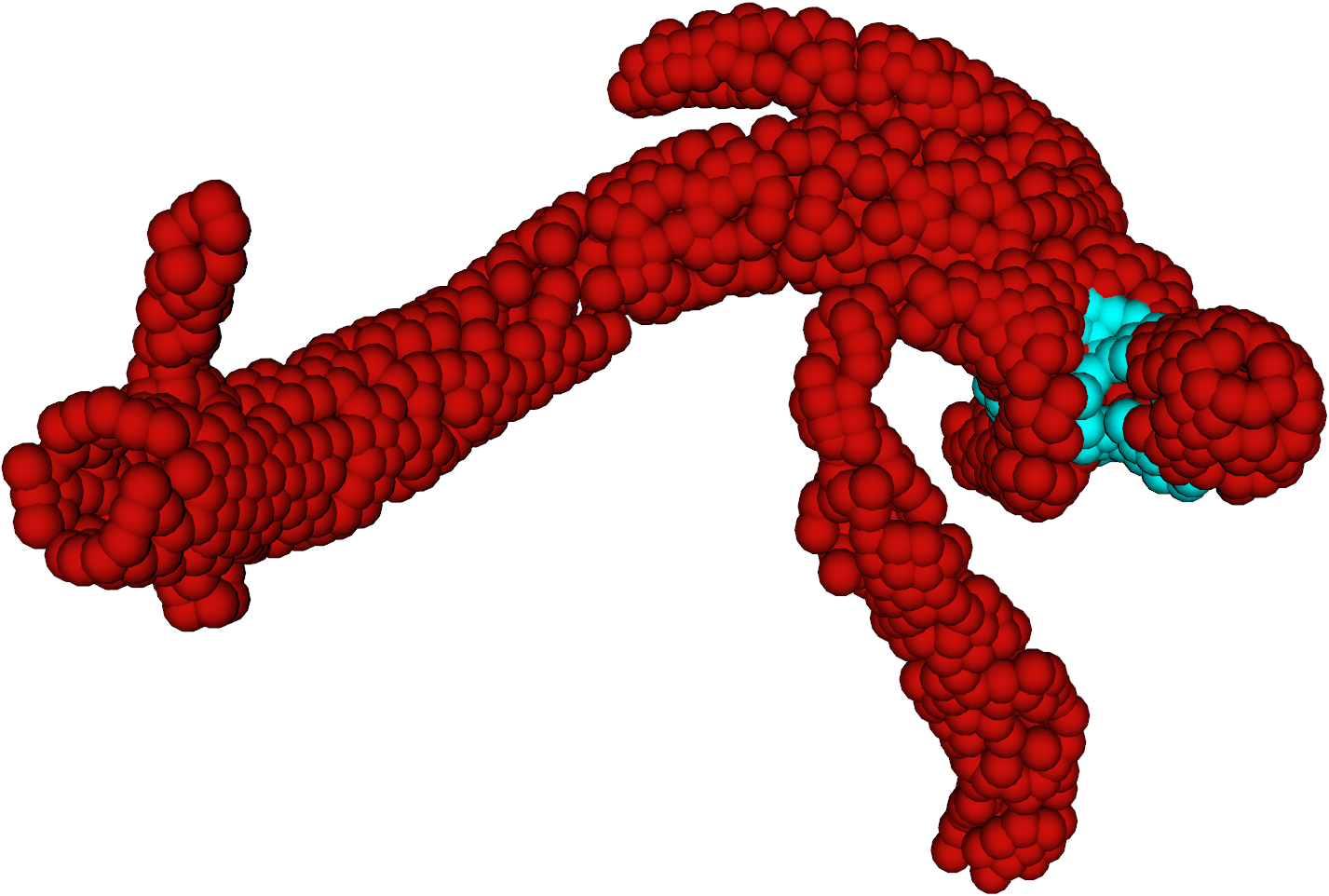}
\end{subfigure}
\begin{subfigure}{.115\textwidth}
  \centering
  \includegraphics[width=0.9\linewidth]{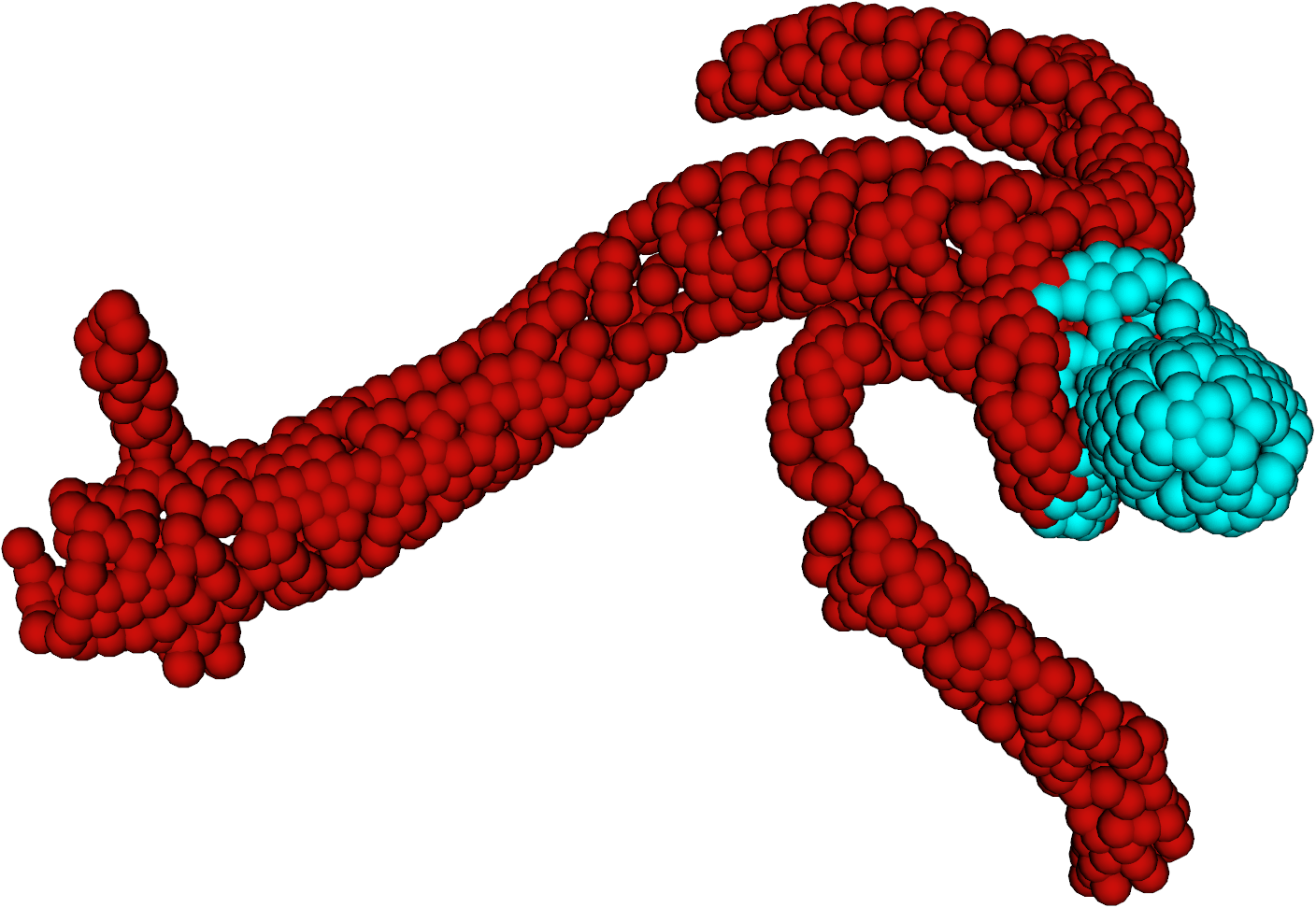}
\end{subfigure}
\begin{subfigure}{.115\textwidth}
  \centering
  \includegraphics[width=0.9\linewidth]{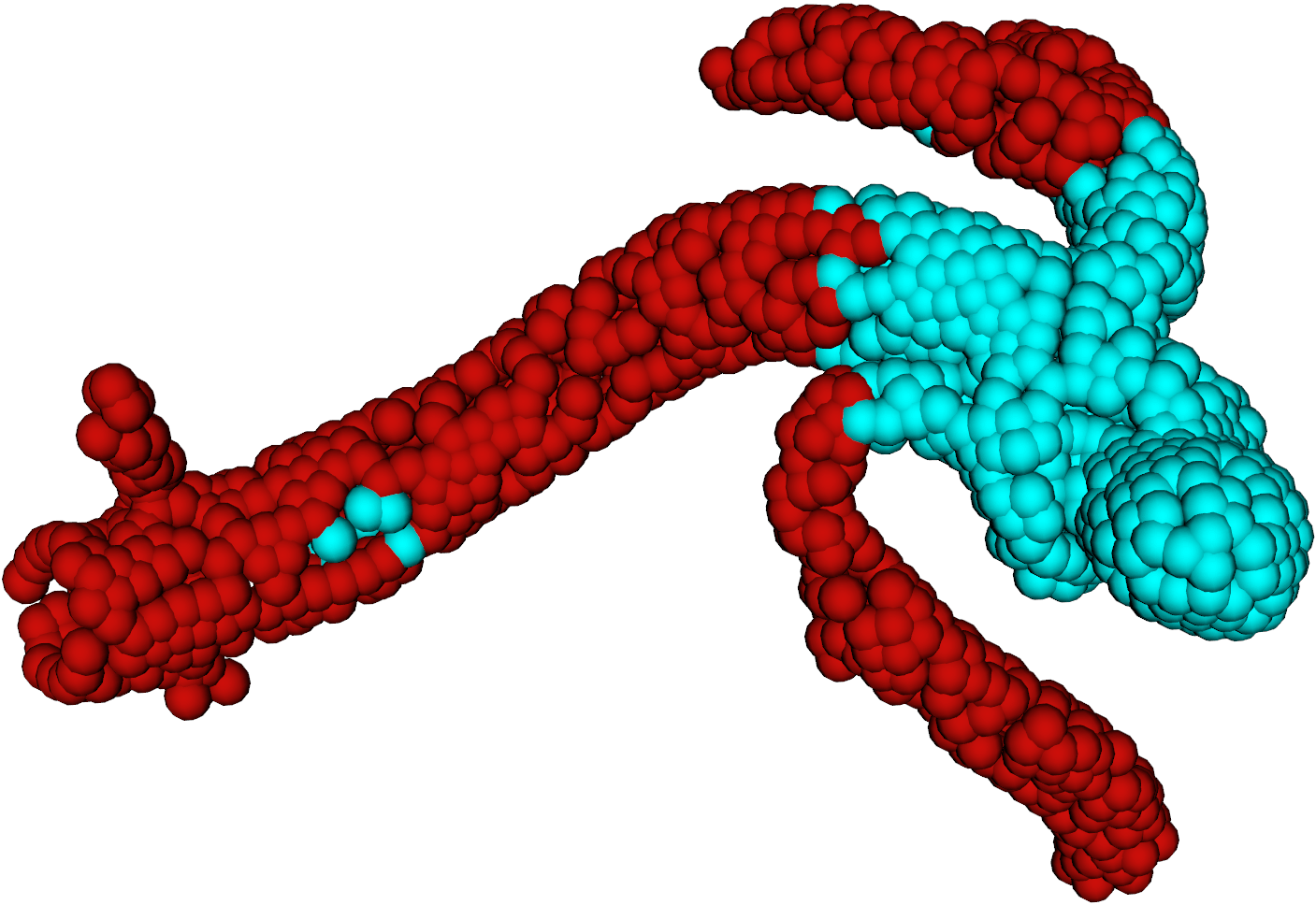}
\end{subfigure}

\vspace{5pt}

\begin{subfigure}{.115\textwidth}
  \centering
  \includegraphics[width=0.9\linewidth]{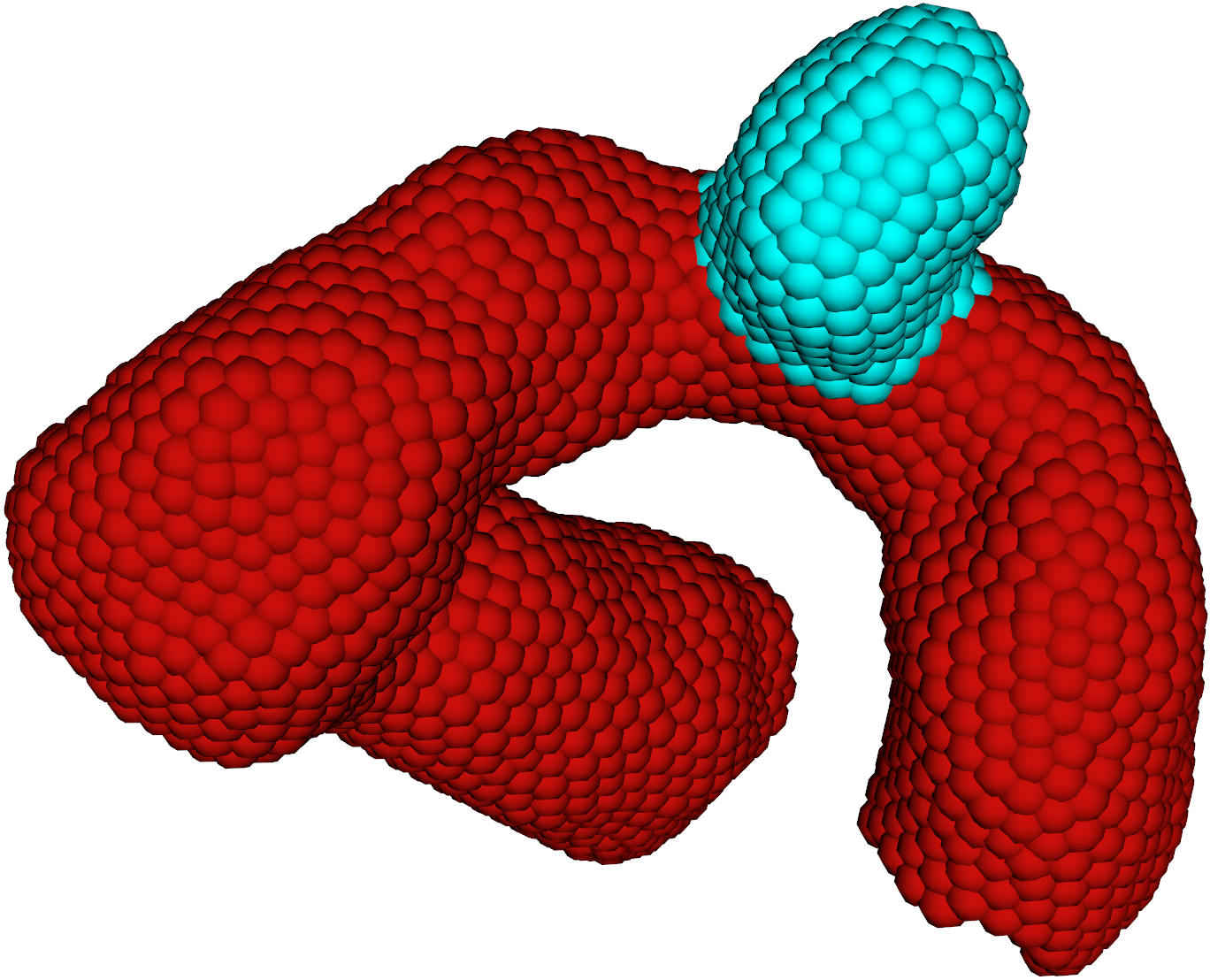}
\end{subfigure}%
\begin{subfigure}{.115\textwidth}
  \centering
  \includegraphics[width=0.9\linewidth]{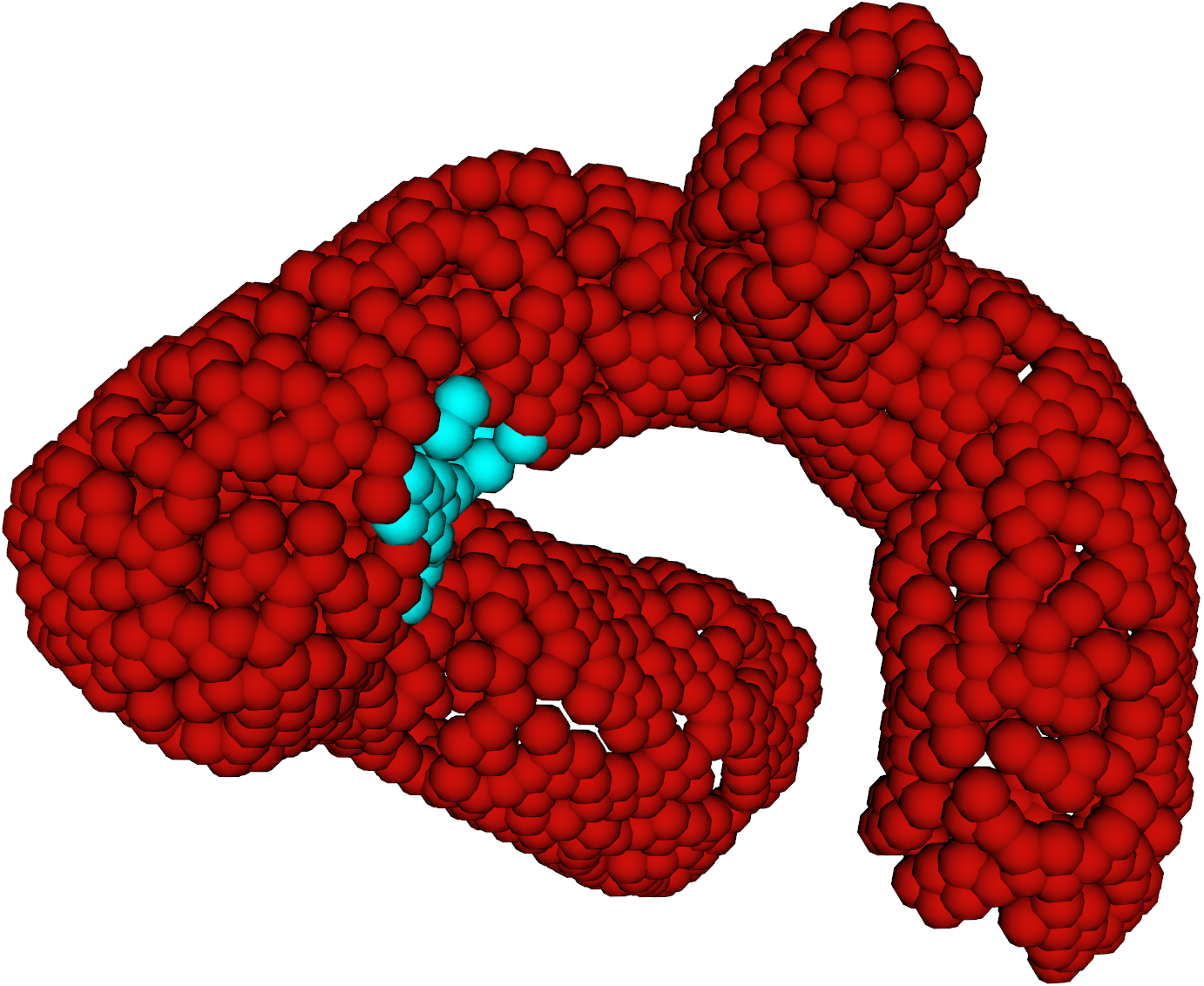}
\end{subfigure}
\begin{subfigure}{.115\textwidth}
  \centering
  \includegraphics[width=0.9\linewidth]{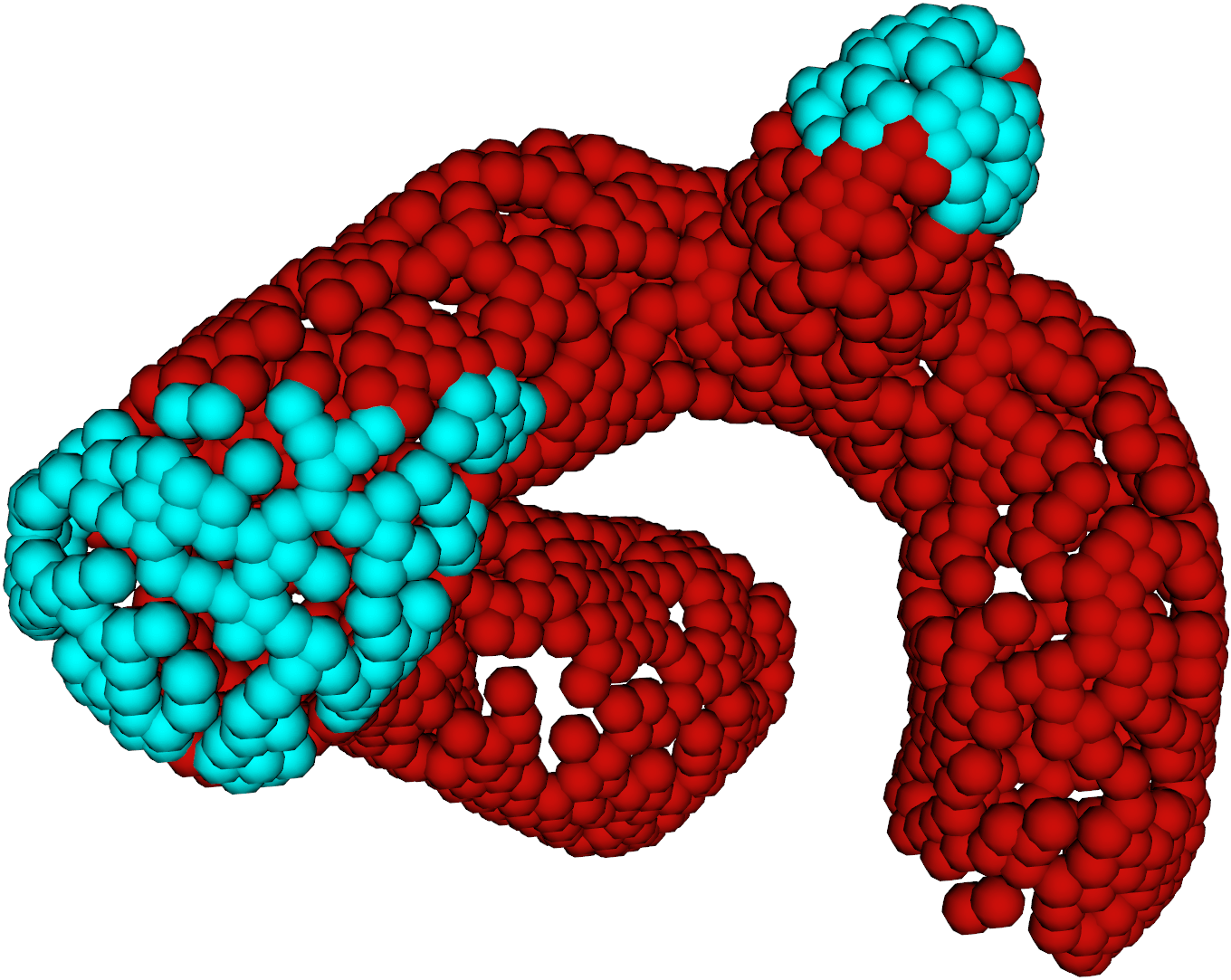}
\end{subfigure}
\begin{subfigure}{.115\textwidth}
  \centering
  \includegraphics[width=0.9\linewidth]{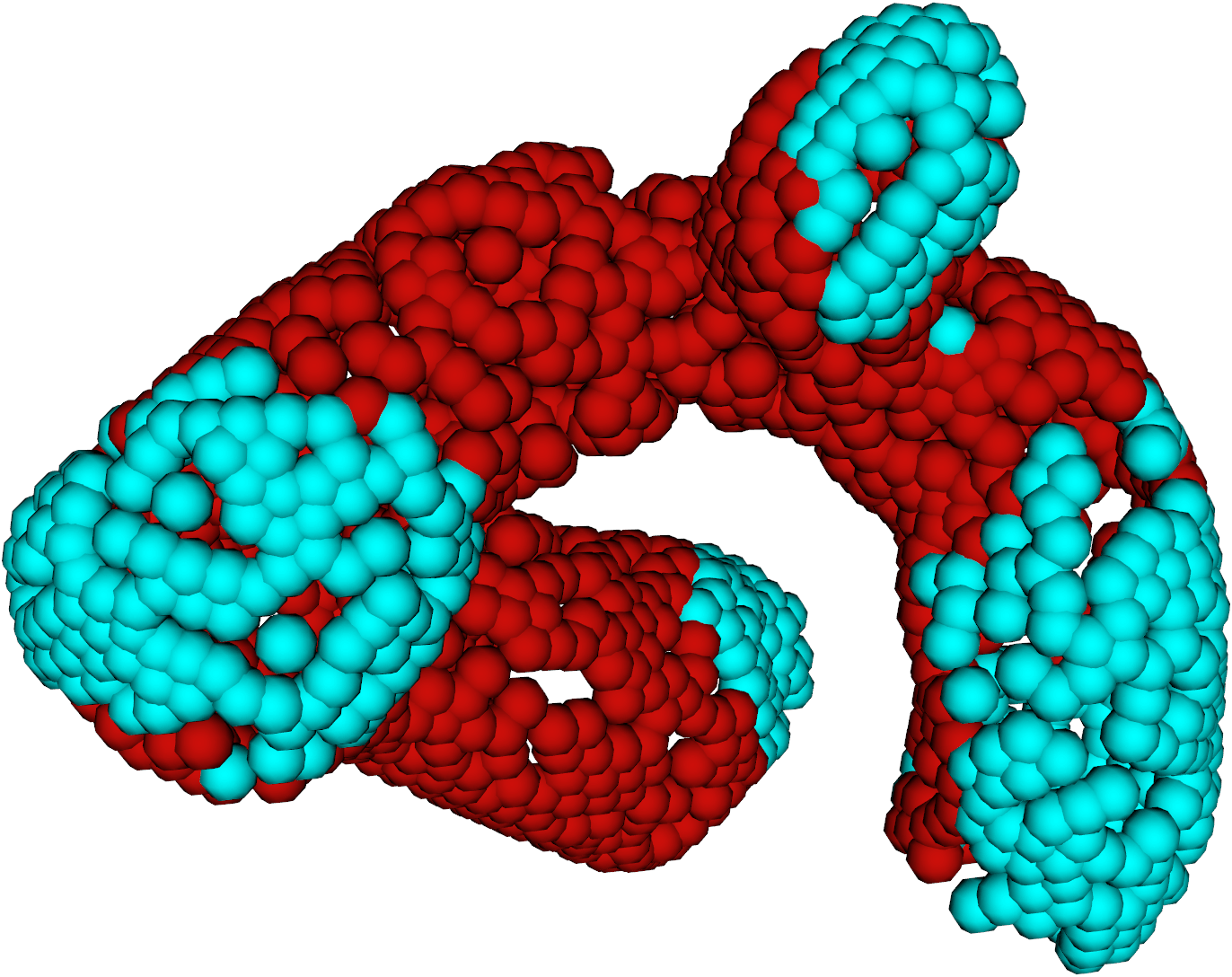}
\end{subfigure}

\vspace{2pt}

\begin{subfigure}{.115\textwidth}
  \centering
  \includegraphics[width=0.9\linewidth]{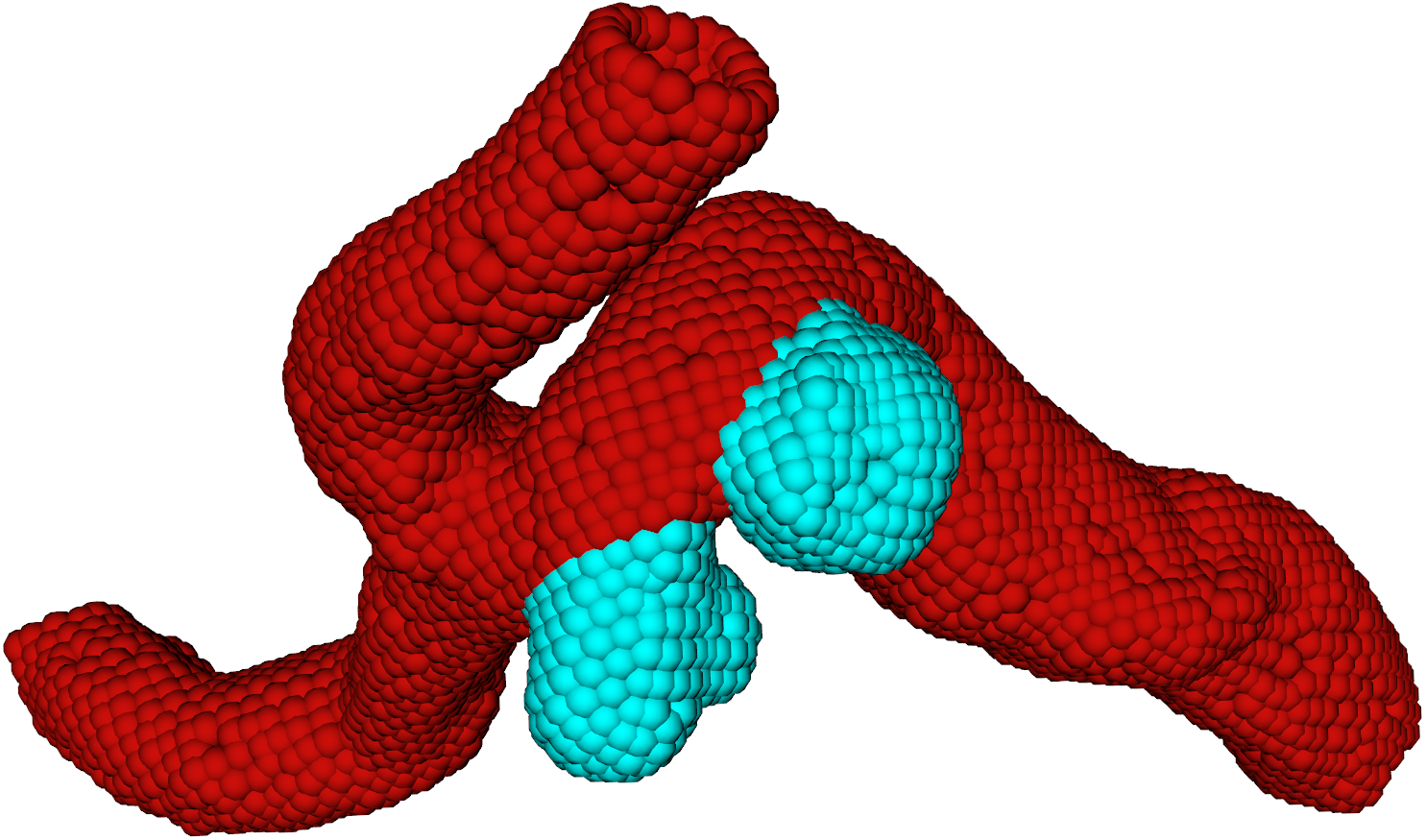}
\end{subfigure}%
\begin{subfigure}{.115\textwidth}
  \centering
  \includegraphics[width=0.9\linewidth]{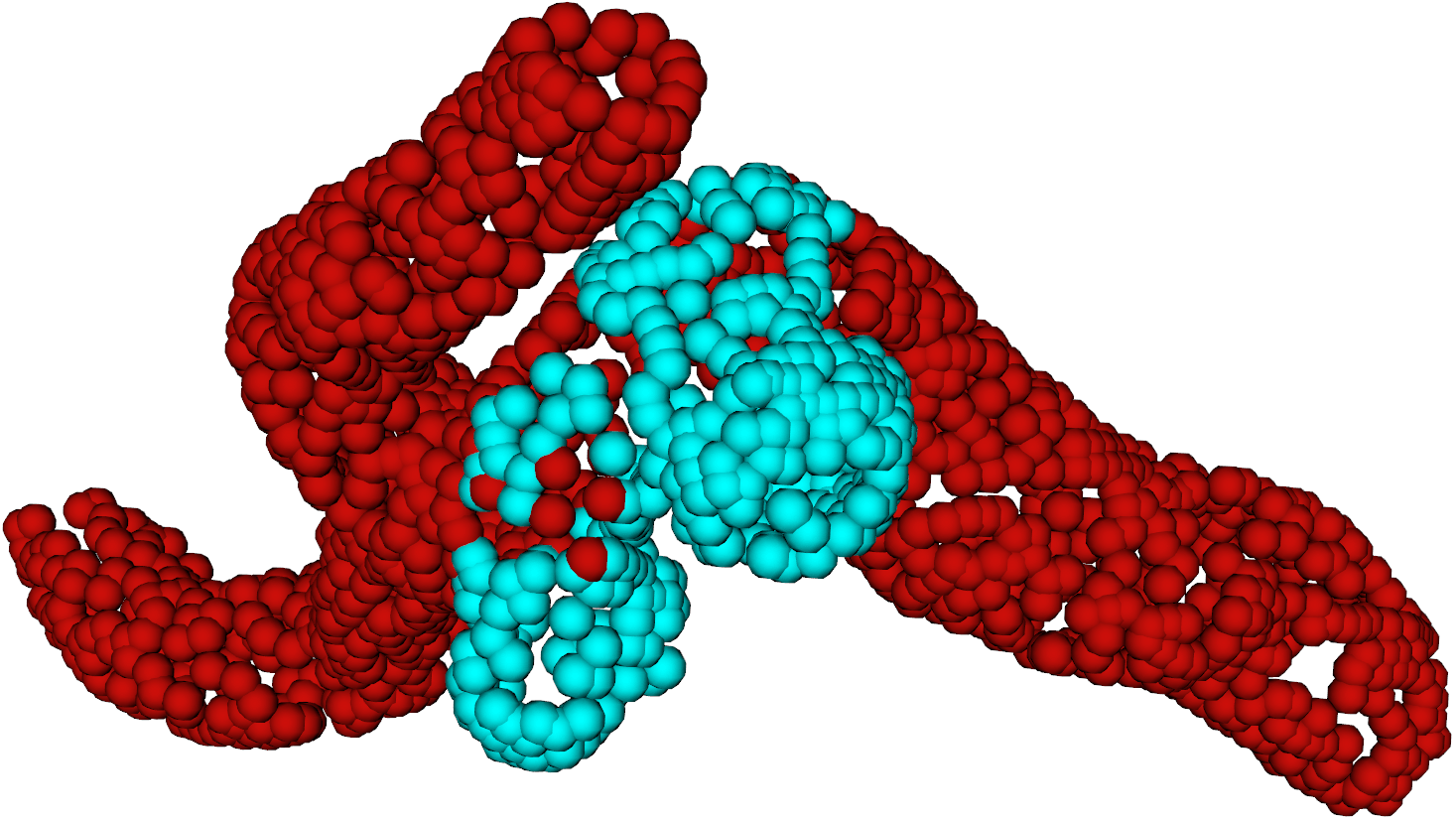}
\end{subfigure}
\begin{subfigure}{.115\textwidth}
  \centering
  \includegraphics[width=0.9\linewidth]{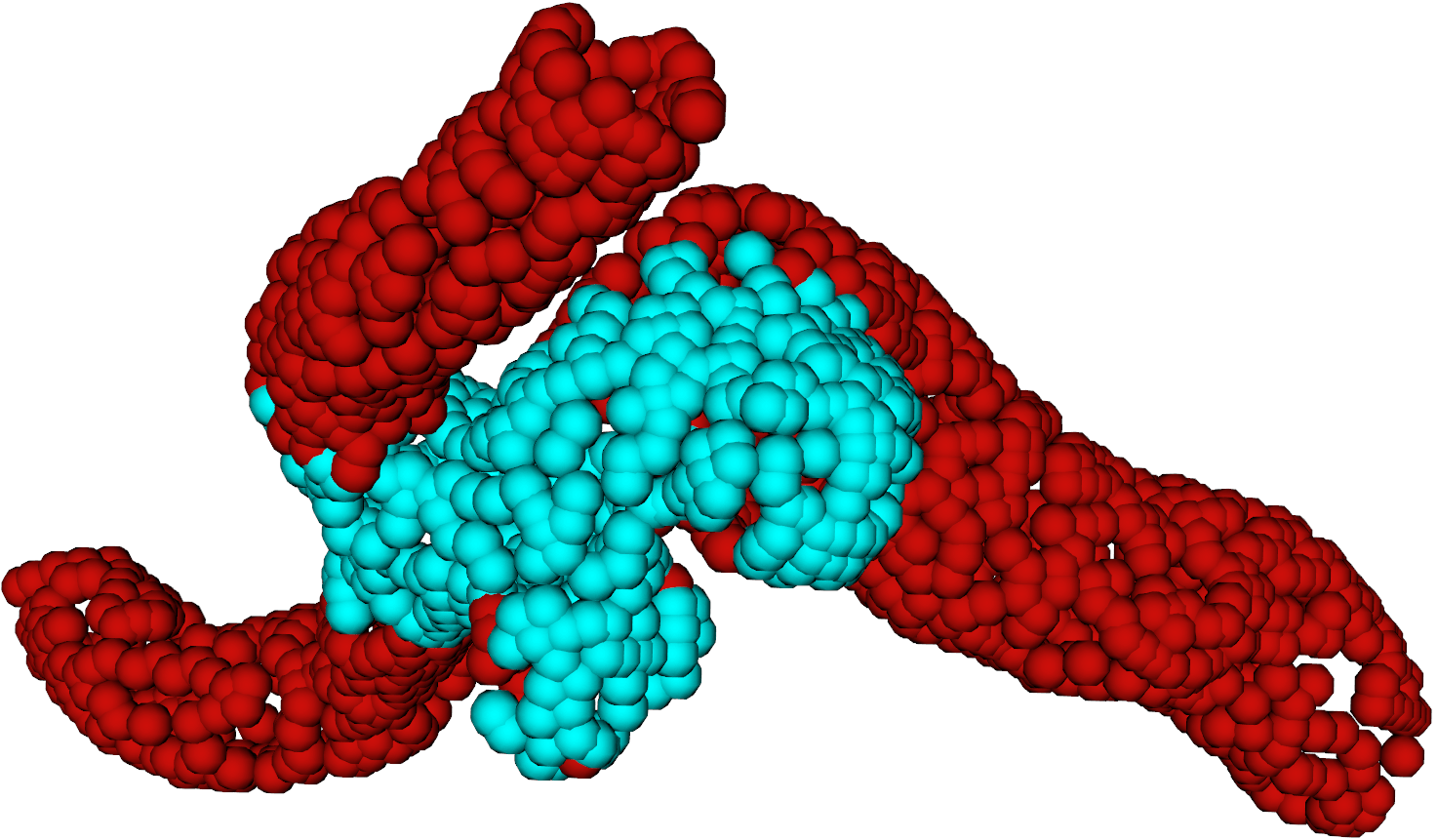}
\end{subfigure}
\begin{subfigure}{.115\textwidth}
  \centering
  \includegraphics[width=0.9\linewidth]{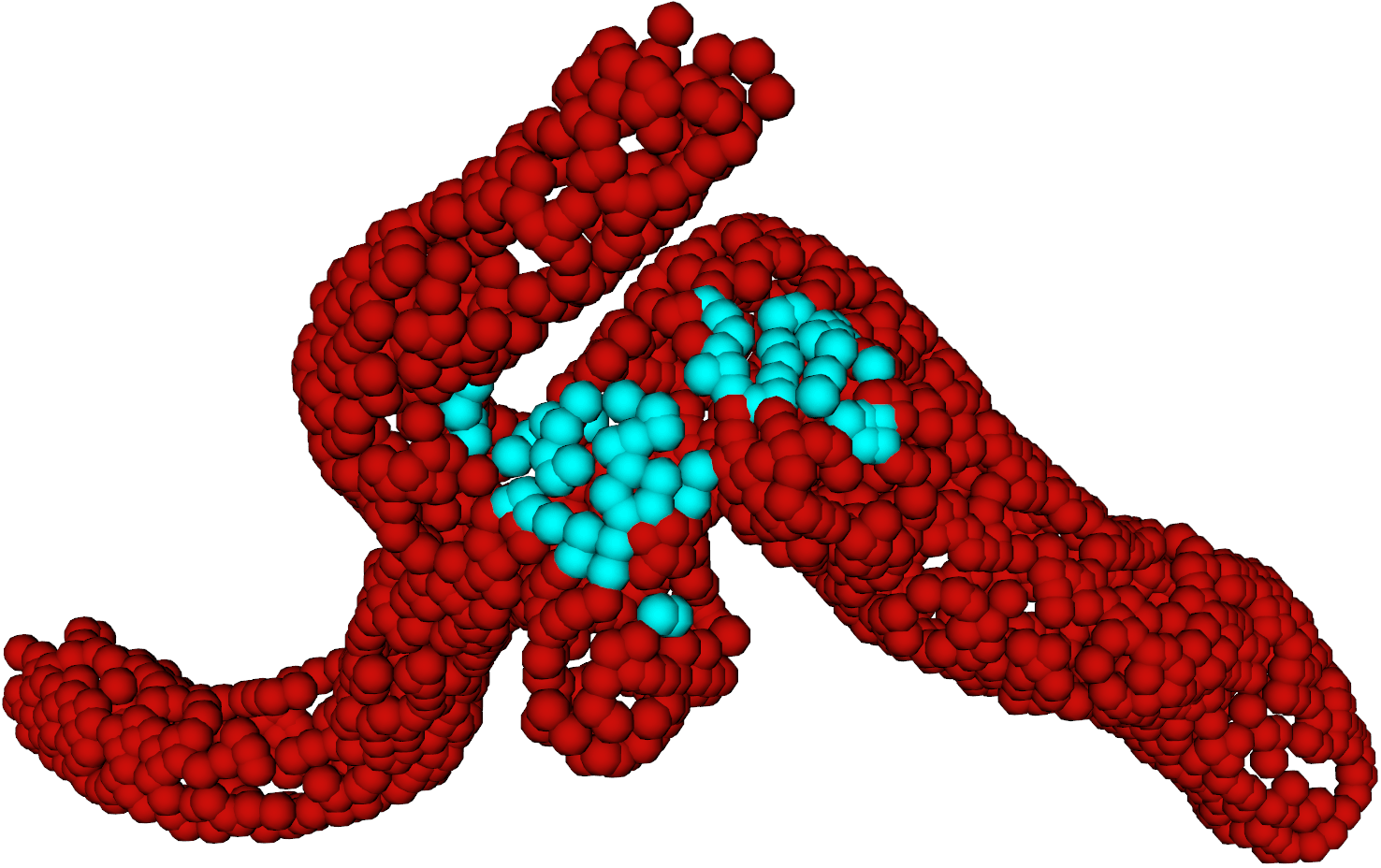}
\end{subfigure}

\vspace{2pt}

\begin{subfigure}{.115\textwidth}
  \centering
  \includegraphics[width=0.9\linewidth]{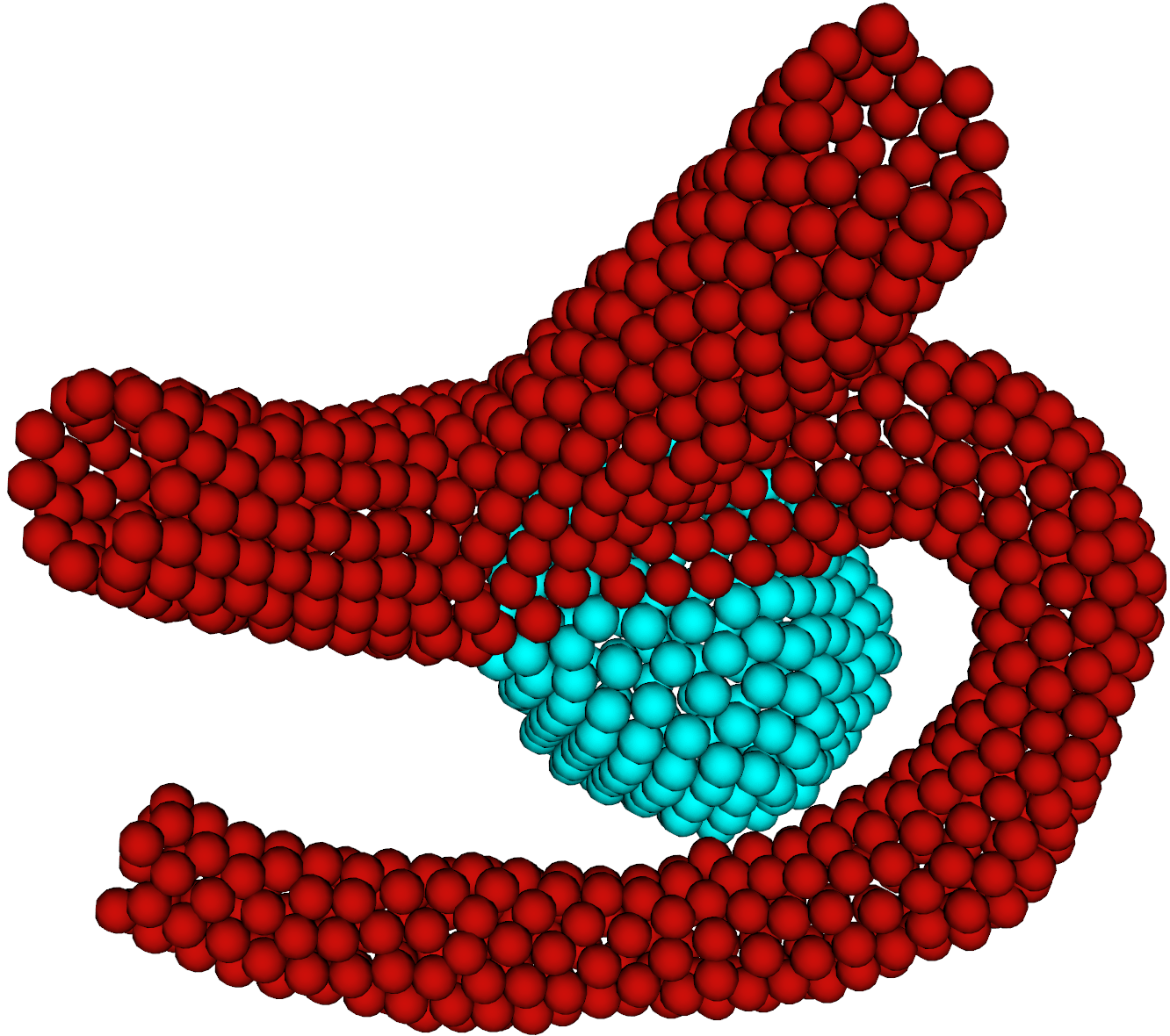}
\end{subfigure}%
\begin{subfigure}{.115\textwidth}
  \centering
  \includegraphics[width=0.85\linewidth]{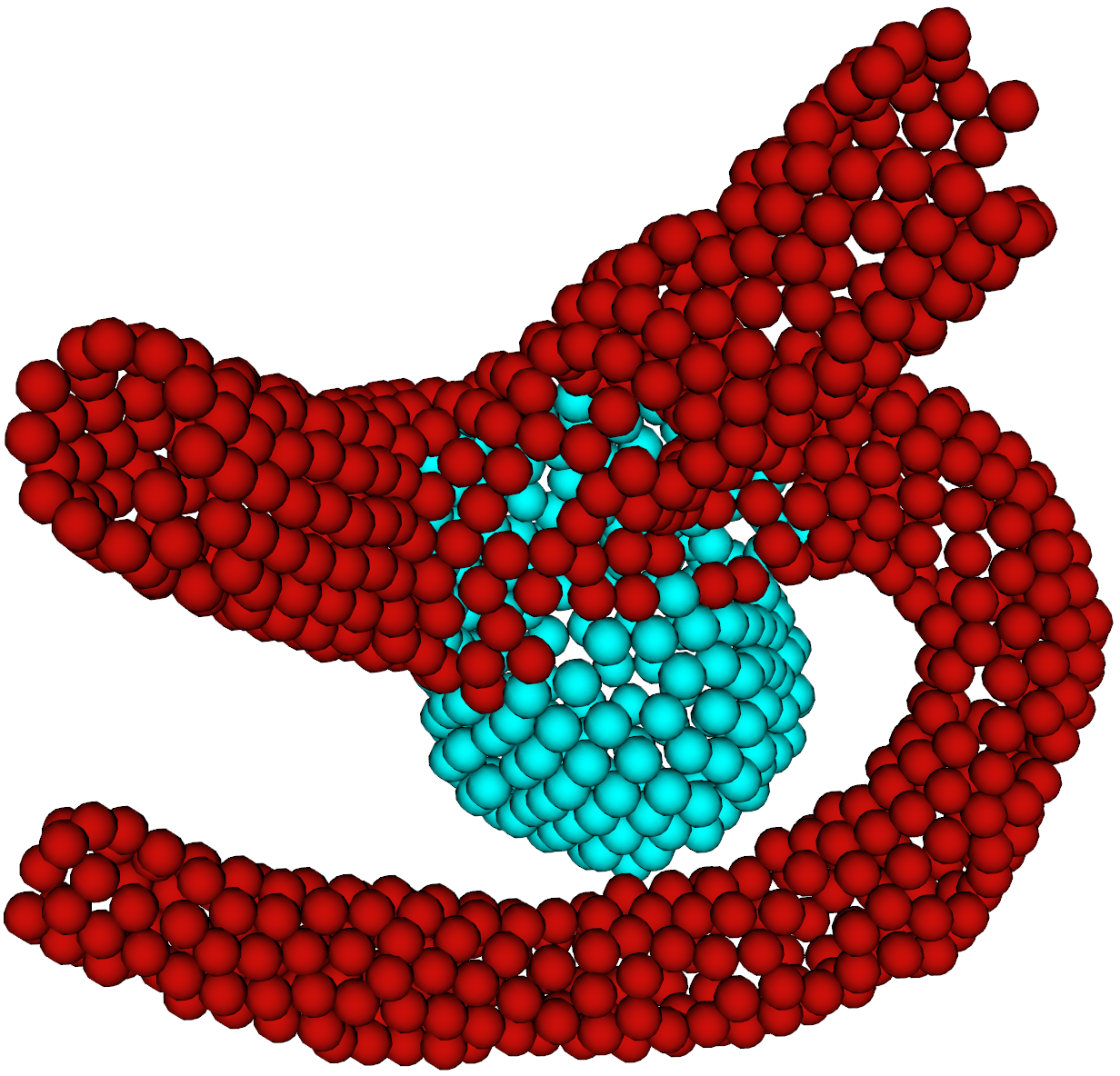}
\end{subfigure}
\begin{subfigure}{.115\textwidth}
  \centering
  \includegraphics[width=0.9\linewidth]{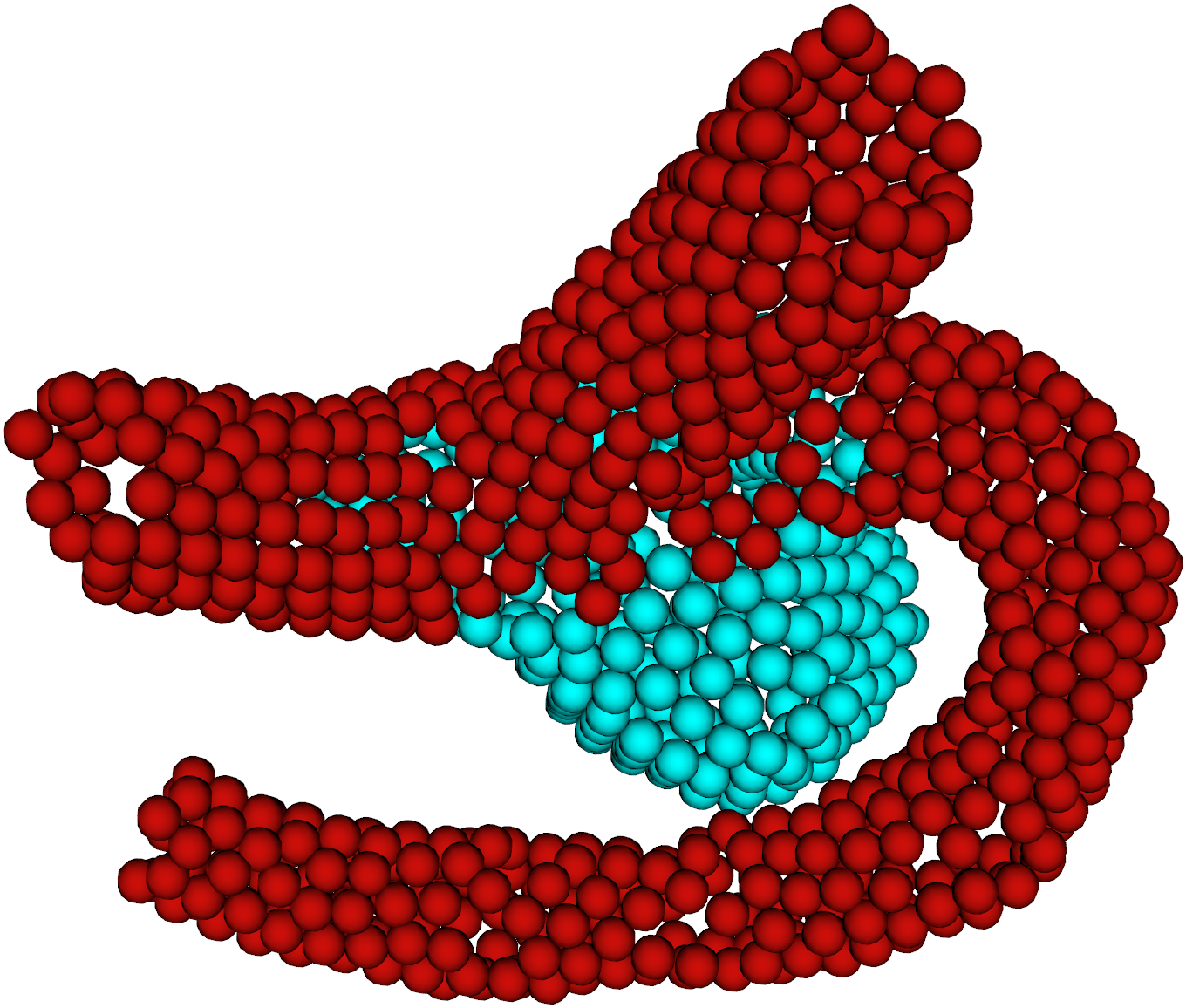}
\end{subfigure}
\begin{subfigure}{.115\textwidth}
  \centering
  \includegraphics[width=0.9\linewidth]{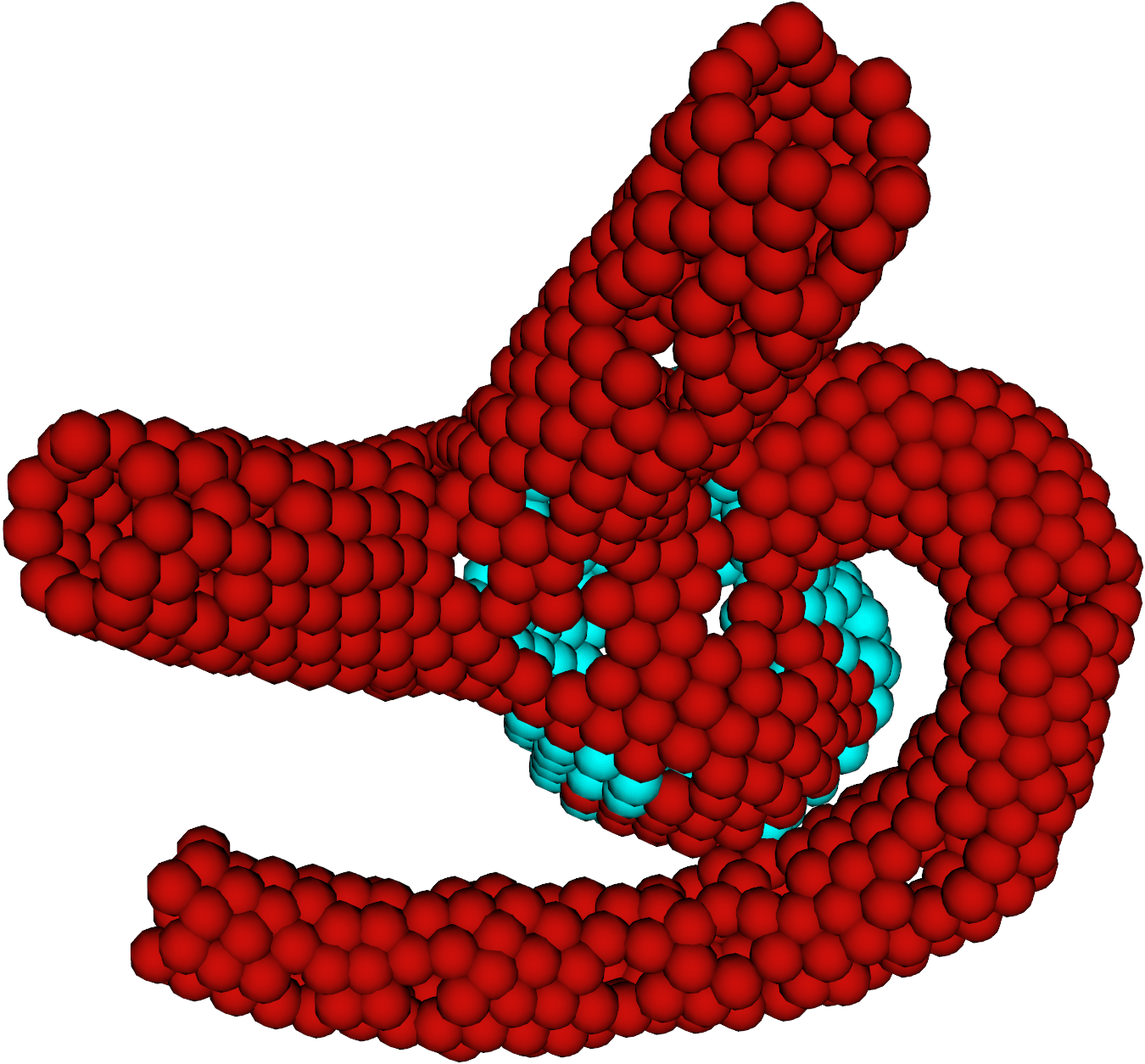}
\end{subfigure}

\vspace{8pt}
\hspace{12pt} Goundtruth ~~~~~ PointConv ~~~~~ PN++g ~~~~~~~~~ PN++
\caption{Results comparison of three networks. More details are described in the supplementary material.} 
\label{fig:discussion}
\end{figure}

\textbf{Methods based on points.}
The segmentation methods based on point cloud obtained good results and maintained the same level with the results on ShapeNet~\cite{shapenet2015}. SO-Net showed excellent performance on IOU and DSC of aneurysms, while PointConv had the best result on parent blood vessels.
PN++ had the third-best performance and had the fastest training speed (5s per epoch, and converged at approximately an epoch of 115 on GTX 1080 Ti). Meanwhile, PointCNN had the slowest training speed (24s per epoch, and converged at approximately an epoch of 500 on GTX 1080 Ti) and moderate segmentation accuracy. SpiderCNN did not have the same performance as it had on the ShapeNet, but CI95 was unusually high.
Besides the methods mentioned in Section~\ref{sec:seg}, 
we also tried 3D CapsuleNet~\cite{zhao20193d}, but it classified every point into the healthy blood vessel,
which shows its limited generalization crossing datasets.

\textbf{Resolution of voxels.}
Methods based on voxels achieved relatively low IOU and DSC on each fold. The performance of SSCN grew as the resolution was increased from 24 to 40 (the resolution 24 was offered by the authur in the code). But the average IOU had a fluctuation of about 8\%, which was quite obvious compared to other methods (about 2\%).
Based on the paper of PointGrid, $N=16$ and $K=2$ were recommended parameters. However, we noticed that the combination of $N=32$ and $K=2$ achieved the highest scores.

\textbf{Common poorly segmented 3D models.}
Most models were segmented excellent as top two rows in Figure~\ref{fig:discussion}.
However, the accuracy dropped when the aneurysm occupied a small size ratio of the segment, like the third and fourth row. Meanwhile, the segmentation performance of aneurysms with a large size ratio was satisfactory. 
The fifth row shows a special segment with 2 aneurysms. Although most of methods failed to segment it, PointConv and PN++ with geodesic information maintained a good performance. 

\textbf{Geodesic information.}
Compared to other CAD model datasets, the complex shapes of blood vessels is a different challenge in part segmentation. Methods based on points usually use the Euclidean distance to estimate the relevance between points. However, it is not ideal for our dataset. For example, PN++ misclassified the aneurysm points close to the blood vessels even with normal information, as shown in the last row of Figure~\ref{fig:discussion}. While, by using geodesic distance, PN++ learned more exact spatial structure. PointConv also segmented it well. Its excellent performance can be attributed to the network learning the parameters of spatial filters.
In addition, MeshCNN segmented every aneurysm decently although the overall performance is not best, which owes to its convolution on meshes providing information on manifolds.

\renewcommand{\arraystretch}{1.15}
\begin{table*}[th]
\begin{center}
\setlength\tabcolsep{3pt} 
\begin{tabular}{ccccccccccc}
\hline
\\ [-2.5ex]
\multicolumn{2}{c}{\multirow{2}{*}{Network}} & \multirow{2}{*}{Input} & \multicolumn{2}{c}{IoU ($\%$)} & \multicolumn{2}{c}{CI 95 ($\%$)} & \multicolumn{2}{c}{DSC ($\%$)} & \multicolumn{2}{c}{CI 95 ($\%$)} \\
& & & V. & A. & V. & A. & V. & A. & V. & A. \\ 
\\ [-2.5ex]
\hline 
\\ [-2.3ex]

\multirow{3}{*}{\rotatebox{90}{Point}} & \multirow{3}{*}{SO-Net}
& 512 & $94.22$ & $80.14$ & $92.12\sim{96.32} $ & $73.71\sim{86.57}$ & $96.95$ & $87.90$ & $95.79\sim{98.12}$ & $83.14\sim{92.66}$ \\
& & 1024 & $94.42$ & $80.99$ & $92.39\sim{96.45}$ & $74.71\sim{87.27}$ & $97.06$ & $88.41$ & $95.93\sim{98.19}$ & $83.65\sim{93.17}$ \\
& & 2048 & $94.46$ & $\mathbf{81.40}$ & $92.51\sim{96.41}$ & $\mathbf{75.37}\sim{87.43}$ & $97.09$ & $\mathbf{88.76}$ & $96.02\sim{98.16}$ & $\mathbf{84.38}\sim{93.15}$ \\ 
\\ [-2.3ex]
\hline
\\ [-2.3ex]
\multirow{3}{*}{\rotatebox{90}{Point}} & \multirow{3}{*}{PointConv}
& 512 & $94.16$ & $79.09$ & $91.76\sim{96.56} $ & $70.26\sim{87.92}$ & $96.89$ & $86.01$ & $95.55\sim{98.24}$ & $78.34\sim{93.69}$ \\
& & 1024 & $94.59$ & $79.42$ & $92.53\sim{96.66}$ & $70.55\sim{\mathbf{88.29}}$ & $97.15$ & $86.29$ & $96.00\sim{98.30}$ & $78.33\sim{\mathbf{94.25}}$ \\
& & 2048 & $\mathbf{94.65}$ & $79.53$ & $\mathbf{92.64}\sim{\mathbf{96.67}}$ & $70.96\sim{88.10}$ & $\mathbf{97.18}$ & $86.52$ & $\mathbf{96.06}\sim{\mathbf{98.30}}$ & $78.95\sim{94.09}$ \\
\\ [-2.3ex]
\hline
\\ [-2.3ex]
\multirow{3}{*}{\rotatebox{90}{Point}} & \multirow{3}{*}{PN++g}
& 512 & $93.34$ & $75.74$ & $91.07\sim{95.60}$ & $66.51\sim{84.97}$ & $96.47$ & $83.90$ & $95.19\sim{97.74}$ & $75.68\sim{92.12}$ \\
& & 1024 & $93.28$ & $76.53$ & $90.93\sim{95.62}$ & $67.91\sim{85.15}$ & $96.43$ & $84.82$ & $95.12\sim{97.75}$ & $77.65\sim{91.99}$ \\
& & 2048 & $93.60$ & $76.95$ & $91.44\sim{95.75}$ & $68.76\sim{85.14}$ & $96.62$ & $85.18$ & $95.41\sim{97.82}$ & $78.39\sim{91.98}$ \\ 
\\ [-2.3ex]
\hline
\\ [-2.3ex]
\multirow{3}{*}{\rotatebox{90}{Point}} & \multirow{3}{*}{PN++}
& 512 & $93.42$ & $76.22$ & $90.91\sim{95.92}$ & $66.70\sim{85.73}$ & $96.48$ & $83.92$ & $95.04\sim{97.92}$ & $75.46\sim{92.38}$\\
& & 1024 & $93.35$ & $76.38$ & $91.10\sim{95.60}$ & $67.96\sim{84.80}$ & $96.47$ & $84.62$ & $95.20\sim{97.74}$ & $77.45\sim{91.78}$\\
& & 2048 & $93.24$ & $76.21$ & $90.93\sim{95.56}$ & $67.99\sim{84.43}$ & $96.40$ & $84.64$ & $95.08\sim{97.72}$ & $77.71\sim{91.57}$ \\ 
\\ [-2.3ex]
\hline
\\ [-2.3ex]
\multirow{3}{*}{\rotatebox{90}{Point}} & \multirow{3}{*}{PointCNN}
& 512 & $92.49$ & $70.65$ & $89.77\sim{95.22}$ & $58.89\sim{82.42}$ & $95.97$ & $78.55$ & $94.41\sim{97.54}$ & $67.37\sim{89.73}$ \\
& & 1024 & $93.47$ & $74.11$ & $91.11\sim{95.84}$ & $63.54\sim{84.68}$ & $96.53$ & $81.74$ & $95.20\sim{97.86}$ & $71.88\sim{91.59}$ \\
& & 2048 & $93.59$ & $73.58$ & $91.45\sim{95.73}$ & $62.81\sim{84.35}$ & $96.62$ & $81.36$ & $95.43\sim{97.80}$ & $71.39\sim{91.33}$ \\ 
\\ [-2.3ex]
\hline
\\ [-2.3ex]
\multirow{3}{*}{\rotatebox{90}{Mesh}} & \multirow{3}{*}{MeshCNN}
& 750 & $85.43$ & $55.63$ & $81.22\sim{89.64}$ & $46.53\sim{64.73}$ & $91.71$ & $68.65$ & $88.86\sim{94.56}$ & $60.16\sim{77.14}$\\
& & 1500 & $90.86$ & $71.32$ & $88.20\sim{93.53}$ & $65.01\sim{77.62}$ & $95.10$ & $82.21$ & $93.55\sim{96.64}$ & $77.26\sim{87.16}$\\ 
& & 2250 & $90.34$ & $71.60$ & $87.34\sim{93.34}$ & $63.99\sim{79.21}$ & $94.77$ & $81.87$ & $93.01\sim{96.53}$ & $75.72\sim{88.01}$\\ 
\\ [-2.3ex]
\hline
\\ [-2.3ex]
\multirow{3}{*}{\rotatebox{90}{Point}}& \multirow{3}{*}{SpiderCNN}
& 512 & $90.16$ & $67.25$ & $86.34\sim{93.98}$ & $55.21\sim{79.29}$ & $94.53$ & $75.82$ & $92.17\sim{96.88}$ & $64.07\sim{87.56}$ \\
& & 1024 & $87.95$ & $61.60$ & $83.65\sim{92.24}$ & $48.97\sim{74.23}$ & $93.24$ & $71.08$ & $90.56\sim{95.93}$ & $58.50\sim{83.67}$ \\
& & 2048 & $87.02$ & $58.32$ & $82.57\sim{91.47}$ & $45.21\sim{71.44}$ & $92.71$ & $67.74$ & $89.94\sim{95.47}$ & $54.50\sim{80.98}$ \\ 
\\ [-2.3ex]
\hline
\\ [-2.3ex]
\multirow{3}{*}{\rotatebox{90}{Voxel}} & \multirow{3}{*}{SSCN-F}
& 24 & $87.95$ & $56.56$ & $84.32\sim{91.57}$ & $43.29\sim{69.84}$ & $93.35$ & $66.04$ & $91.14\sim{95.56}$ & $52.28\sim{79.80}$ \\
& & 32 & $88.08$ & $55.26$ & $84.62\sim{91.54}$ & $41.39\sim{69.13}$ & $93.44$ & $64.05$ & $91.36\sim{95.53}$ & $49.45\sim{78.66}$ \\
& & 40 & $90.09$ & $61.45$ & $87.00\sim{93.17}$ & $48.54\sim{74.37}$ & $94.62$ & $70.54$ & $92.78\sim{96.46}$ & $57.16\sim{83.91}$ \\ 
\\ [-2.3ex]
\hline
\\ [-2.3ex]
\multirow{3}{*}{\rotatebox{90}{Voxel}} & \multirow{3}{*}{SSCN-U}
& 24 & $87.43$ & $55.78$ & $83.48\sim{91.37}$ & $42.07\sim{69.48}$ & $93.00$ & $65.03$ & $90.58\sim{95.43}$ & $50.91\sim{79.16}$ \\ 
& & 32 & $86.13$ & $53.52$ & $82.12\sim{90.13}$ & $40.64\sim{66.40}$ & $92.24$ & $64.01$ & $89.72\sim{94.76}$ & $50.95\sim{77.06}$ \\ 
& & 40 & $88.66$ & $57.94$ & $85.25\sim{92.07}$ & $44.92\sim{70.96}$ & $93.78$ & $67.39$ & $91.73\sim{95.83}$ & $54.09\sim{80.64}$ \\ 
\\ [-2.3ex]
\hline
\\ [-2.3ex]
\multirow{3}{*}{\rotatebox{90}{Voxel}} & \multirow{3}{*}{PointGrid}
& 16/2 & $78.32$ & $35.82$ & $73.53\sim{83.12}$ & $25.22\sim{46.42}$ & $87.36$ & $47.33$ & $84.12\sim{90.60}$ & $35.28\sim{59.38}$ \\
& & 16/4 & $79.49$ & $38.23$ & $74.54\sim{84.44}$ & $26.29\sim{50.17}$ & $88.08$ & $49.14$ & $84.73\sim{91.43}$ & $35.65\sim{62.64}$ \\
& & 32/2 & $80.11$ & $42.42$ & $75.60\sim{84.62}$ & $30.51\sim{54.34}$ & $88.50$ & $53.52$ & $85.54\sim{91.45}$ & $40.41\sim{66.63}$  \\ 
\\ [-2.3ex]
\hline
\\ [-2.3ex]
\multirow{3}{*}{\rotatebox{90}{Point}} & \multirow{3}{*}{PointNet}
& 512 & $73.99$ & $37.30$ & $67.43\sim{80.56}$ & $26.17\sim{48.44}$ & $84.05$ & $48.96$ & $79.31\sim{88.79}$ & $36.53\sim{61.38}$ \\
& & 1024 & $75.23$ & $37.07$ & $69.10\sim{81.36}$ & $25.66\sim{48.48}$ & $85.00$ & $48.38$ & $80.69\sim{89.31}$ & $35.63\sim{61.13}$ \\
& & 2048 & $74.22$ & $37.75$ & $67.85\sim{80.60}$ & $26.85\sim{48.64}$ & $84.17$ & $49.59$ & $79.56\sim{88.78}$ & $37.48\sim{61.70}$ \\ 
\\ [-2.3ex]
\hline

\end{tabular}
\caption{Segmentation results of each network. The second column shows the number of input points, edges, or resolutions for the methods based on points, mesh, and voxel, respectively. For PointGrid, the additional refers to the parameter $K$ in the paper. The healthy vessel part and aneurysm part are noted as V. and A., respectively. CI$95$ indicated $95\%$ confidence interval of IoU or DSC. The results are calculated by the mean value of each fold.}
\label{tab:s-results}
\end{center}
\end{table*}

\section{Conclusion and Further Work}

In this paper, we introduced a 3D dataset of intracranial aneurysm with annotation by experts for geometric deep learning networks. The developed tools and data processing pipeline is also released. 
Furthermore, we evaluated and analyzed the state-of-the-art methods of 3D object classification and part segmentation on our dataset. The existing methods are likely to be less effective on complex objects, though they perform well on the segmentation of common ones. 
It is possible to improve further the performance and generalization of networks when geodesic or connectivity information on 3D surfaces is accessible.
The introduction of our dataset can be instructive to the development of new structures of geometric deep learning methods for medical datasets. 

In further work, we will keep increasing processed real data for our dataset. Besides, we will verify the feasibility of synthetic data for data augmentation, which can significantly improve the efficiency of data collection. We hope more deep learning networks will be applied to medical practice.

\section*{Acknowledgements}
This research was supported by AMED under Grant Number JP18he1602001.

\clearpage
\clearpage

{\small
\bibliographystyle{ieee_fullname}
\bibliography{egbib}
}

\end{document}